# Phase diagrams and polar properties of ferroelectric nanotubes and nanowires.

Anna N. Morozovska[*], Maya D. Glinchuk, Eugene A. Eliseev,

Institute for Problems of Materials Science, NAS of Ukraine,

Krjijanovskogo 3, 03142 Kiev, Ukraine,

[*]morozo@i.com.ua, glin@materials.kiev.ua

In this paper we study the size effects of the ferroelectric nanotube and nanowire phase diagrams and polar properties allowing for radial stress and depolarization field influence. The approximate analytical expression for the paraelectric-ferroelectric transition temperature dependence on the radii of nanotube, polarization gradient coefficient, extrapolation length, radial stress (surface tension) and electrostriction coefficient was derived. It was shown that the transition temperature could be higher than the one of the bulk material for negative electrostriction coefficient and small depolarization field. Therefore we predict conservation and enhancement of polar properties in long cylindrical ferroelectric nanoparticles. Obtained results explain the observed ferroelectricity conservation and enhancement in $Pb(Zr,Ti)O_3$ and $BaTiO_3$ nanowires and nanotubes. Moreover, despite made assumptions and approximations our modelling appeared in a surprisingly good agreement with observed ferroelectric and local piezoresponse hysteresis loops.

\* permanent address: V. Lashkaryov Institute of Semiconductor Physics, NAS of Ukraine, 41, pr. Nauki, 03028 Kiev, Ukraine

# I. Introduction

Ferroelectric nanoparticles of different shape are actively studied in nano-physics and nano-technology. The ferroelectric phase was studied in ferroelectric nanowires, nanotubes and nanorods.[1, 2, 3, 4, 5] It is appeared that nanorods and nanotubes posses such polar properties as remnant polarization[1] and piezoelectric hysteresis.[3, 4, 5] Moreover, co-called "confined" geometry does not destroy ferroelectric phase as predicted for spherical particles[6, 7] and observed experimentally [8, 9, 10], but sometimes the noticeable enhancement of ferroelectric properties appears in nano-cylinders.[1, 2, 3, 4, 5, 11]

Yadlovker and Berger[1] reported about the spontaneous polarization enhancement up to 0.25-2$\mu$C/cm$^2$ and ferroelectric phase conservation in Roshelle salt (RS) nanorods (radius 30nm, length 500 nm). Mishina *et al* [11] revealed that ferroelectric phase exists in $PbZr_{0.52}Ti_{0.48}O_3$ (PZT) nanorods with diameter less than 10-20nm. Geneste *et al* [2] studied the size dependence of the ferroelectric properties of $BaTiO_3$ (BT) nanowires from the first principles and showed that the ferroelectric distortion along the wire axis disappears below a critical size of about 1.2nm. The phenomenological description of ferroelectricity enhancement in confined nanorods has been recently proposed.[12][13]

Morrison *et al* [4, 5] demonstrated that $PbZr_{0.52}Ti_{0.48}O_3$ nanotubes (radius $R$=600-700nm, length $50\mu m$) possesses rectangular shape of the piezoelectric hysteresis loop with effective remnant piezoelectric coefficient value compatible with the ones typical for PZT films.[14] Also the authors demonstrated that the ferroelectric properties of the free $BaTiO_3$ nanotubes are perfect.

Poyato et al. [15] with the help of piezoelectric response force microscopy found that nanotube-patterned ("honeycomb") $BaTiO_3$ film of thickness 200-300 nm reveal ferroelectric properties. The inner diameter of the nanotubes ranged from 50 to 100 nm. Also they demonstrated the existence of local piezoelectric and oriented ferroelectric responses, prior to the application of a dc field, in nanotubes-patterned $BaTiO_3$ thin films on Ti substrates synthesized hydrothermally at 200 °C.

Thus, at the first glance recent experimental results contradict the generally accepted viewpoint that the ferroelectric properties disappear under the system volume decreases below the critical one.[16] Actually the aforementioned facts proved that the shape of nanoparticles essentially influences on the critical volume necessary for the ferroelectricity conservation[2] possibly owing to the different depolarization field and mechanical boundary conditions.[17, 18]

In theoretical papers[6, 19] the special attention was paid to size effects, but depolarization field influence on a nanoparticle was neglected. However, it is well known that depolarization field exists in the majority of confined ferroelectric systems[20] and causes the aforementioned size-induced ferroelectricity disappearance in insulator single-domain films and ellipsoidal particles.[21, 22, 23] Both finite size and depolarization field effects lead to the ferroelectric properties degradation, namely the phase transition temperature in spherical nanoparticles is significantly lower the bulk one for most of the cases.[6, 8, 23, 24]

In our consideration we suppose that a nanoparticle surface is covered with a charged layer consisted of the free carriers adsorbed in the ambient conditions (e.g. air with definite humidity or pores filled with a precursor water solution). For instance a thin water layer condensates on the polar oxide surface in the air with humidity 20-50%.[25] The surface charges screen the surrounding medium from the nanoparticle electric field[16] (the case of non-interacting nanoparticles assembly), but the depolarization field inside the particle is caused by inhomogeneous polarization distribution. Thus one could calculate the depolarization field inside a cylindrical nanoparticle under the short-circuit conditions proposed by Kretschmer and Binder.[21]

To the best of our knowledge the thermodynamical consideration of ferroelectric nanotubes polar properties is absent. For the description of nanotubes and nanowires ferroelectric properties we used the Euler-Lagrange equations, which will be solved by means of a direct variational method.[22] The approximate analytical expression for paraelectric-ferroelectric transition temperature dependence on the nanoparticle sizes, extrapolation length, effective radial stress, polarization gradient and electrostriction coupling coefficients *etc* was derived. Note, that the stress is caused by the particle surface clamping by porous matrix, i.e. it is related to surface energy (surface tension).[26] We obtained, that the possible reason of the polar properties enhancement in confined ferroelectric nanotubes and nanowires is the radial stress coupled with polarization via electrostriction effect under the decrease of depolarization field for long cylindrical nanoparticles.

While the influence of depolarization field is obvious, the role of radial stress can be understood as follows. Despite the radial stress conserves the inversion center, it leads to the short-range forces strengthening in lateral direction (caused by the bond contraction) and their weakening in z-direction (caused by the bond elongation). As a result, the long-range correlations become more pronounced in polar direction in comparison with the short-range forces. Allowing for ferroelectricity cooperative nature[16], the stress stimulates the ferroelectric phase appearance at temperatures higher than the bulk Curie one.

## II. Free energy of cylindrical nanoparticles

Let us consider the ferroelectric cylindrical nanotube of outer radius $R_1$, inner radius $R_2$, height $h$ and polarization $P_Z(\rho, \psi, z)$ oriented along z–axes. The external electric field is $\mathbf{E} = (0, 0, E_0)$ (see Fig.1).

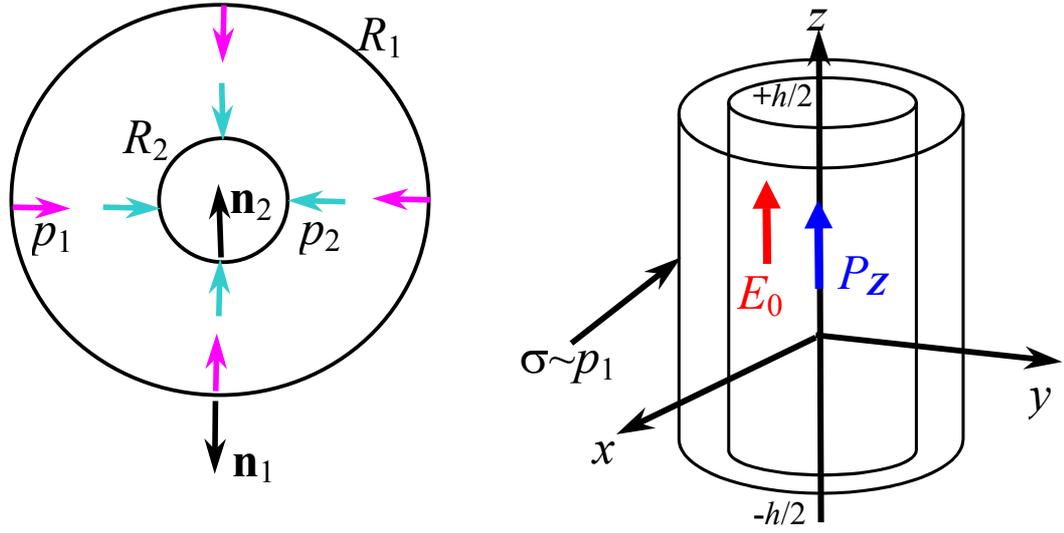

FIG. 1. (Color online). The nanotube under the radial stress.

The Euler-Lagrange equation for the polarization can be obtained by the variation on polarization of the free energy functional $\Delta G = \Delta G_V + \Delta G_S$ consisted from the bulk part $\Delta G_V$ and the surface one $\Delta G_S$ (see e.g. Ref. 18).

The bulk part $\Delta G_V$ acquires the form:

$$\Delta G_V = \int\limits_{-h/2}^{h/2} dz \int\limits_{0}^{2\pi} d\psi \int\limits_{R_2}^{R_1} \rho d\rho \left( \frac{\alpha_R(T)}{2} P_Z^2 + \frac{\beta}{4} P_Z^4 + \frac{\gamma}{6} P_Z^6 + \frac{\delta}{2} (\nabla P_Z)^2 - P_Z \left( E_0 + \frac{E_Z^d}{2} \right) \right) \quad (1)$$

Material coefficients $\delta > 0$ and $\gamma > 0$, while $\beta < 0$ for the first order phase transitions or $\beta > 0$ for the second order ones. The coefficient $\alpha_R(T)$ in Eq.(1) should be renormalized by the external stress (see e.g. Ref. 24, 27). In Appendix A we study the influence of the effective surface stresses on a nanotube sidewalls and derived the expression for $\alpha_R(T)$:

$$\alpha_R(T, R_1, R_2) = \alpha_T (T - T_C) - 2Q_{12}\sigma(R_1, R_2) \quad . \quad (2a)$$

Here parameters $T_C$ and $Q_{12}$ are respectively Curie temperature and electrostriction coefficient of the bulk material, $\alpha_T$ is proportional to the inverse Curie constant. The stress $\sigma(R_1, R_2)$ is caused by the radial pressures $p_1$ and $p_2$ (see Fig.1). We modified the solution of this Lame's problem allowing for stress relaxation as following:

$$\sigma(R_1, R_2) \approx \frac{1}{1 - (R_2/R_1)^2} \left( p_1 + \left( \frac{R_2^2}{R_1^2} \right) p_2 \right) \exp\left( -\frac{\Delta R}{R_1 - R_2} \right) \quad (2b)$$

Parameter $\Delta R$ is the characteristic thickness of a nanotube, below which the factor $\dfrac{1}{1 - (R_2/R_1)^2}$ becomes too high; thus it characterizes the stress exponential relaxation via dislocation appearance[28] and so prevents the strong increase of the stress. For the case when the radial stress is caused by surface energy, we assume that pressures $p_1 = -\mu_1/R_1$ and $p_2 = -\mu_2/R_2$ are related to the surface tension, where $\mu_{1,2} > 0$ are the effective surface tension coefficients between the nanotube and its interface.[10, 29] In particular case $\mu_1 = \mu_2 = \mu$ one obtains that $\sigma(R_1, R_2) = -\dfrac{\mu}{R_1 - R_2}\exp\left(-\dfrac{\Delta R}{R_1 - R_2}\right)$ and so it reaches maximum at $R_1 - R_2 = \Delta R$, thus $\Delta R = \mu/\sigma_{max}$. It will be important that the stress $\sigma(R_1, R_2) \sim \dfrac{1}{R_1 - R_2}$ does not vanishes for large but thin nanotubes (i.e. $R_{1,2} \to \infty$ but $R_1 - R_2 = const$) in contrast to nanowires ($R_2 = 0$), where $\sigma(R_1) \sim \dfrac{1}{R_1}$ disappears at $R_1 \to \infty$.

The exact expression for depolarization field $\mathbf{E}^d(\rho, \psi, z)$ inside the cylindrical nanoparticle covered with screening charges is derived in Appendix B (see Eqs.(B.8) and (B.13)). The field is highest for a single-domain nanotube, namely its estimation for a thick tube ($R_1 - R_2 >> \sqrt{\delta}$) has the form:

$$E_Z^d(\rho, \psi, z) = -\eta\left(P_Z(\rho, \psi, z) - \frac{2}{h}\int_{-h/2}^{h/2} dz P_Z(\rho, \psi, z)\right),$$
$$\eta = \frac{4\pi}{1 + (k_{01}h/2\pi R_1)^2}. \tag{3}$$

Hereinafter $k_{01}(R_1, R_2)$ is the lowest root of the equation $J_0\left(k_{01}\dfrac{R_2}{R_1}\right)N_0(k_{01}) - J_0(k_{01})N_0\left(k_{01}\dfrac{R_2}{R_1}\right) = 0$ ($J_0(x)$ and $N_0(x)$ are Bessel and Neiman functions of zero order respectively). The function $\eta \sim (R_1/h)^2 << 1$ for the prolate tube with $R_1 << h$,[20] whereas $\eta \to 4\pi$ for the oblate one with $R_1 >> h$.[21] It should be noted that the depolarization field is absent outside the particles in the framework of our model. Therefore the interaction of such nanoparticles is practically absent due to the screening. Their composite can be considered as the assembly of independent particles.

The surface part of the polarization-dependent free energy $\Delta G_S$ is supposed proportional to square of polarization on the particle surface $S$, namely $\Delta G_S = \dfrac{\delta}{2}\int_S \dfrac{ds}{\lambda} P_S^2$ ($\lambda$ is the extrapolation length[6, 19]). A nanotube has upper and bottom surfaces $z = h/2$, $z = -h/2$ and sidewalls $\rho = R_{1,2}$, so its surface energy $\Delta G_S$ acquires the form:

$$\Delta G_S = \delta \int_0^{2\pi} d\psi \left(\int_{R_2}^{R_1} \frac{\rho}{\lambda_b} d\rho \left(P_Z^2\left(z = \frac{h}{2}\right) + P_Z^2\left(z = -\frac{h}{2}\right)\right) + \int_{-h/2}^{h/2} dz \left(\frac{R_1}{\lambda_S} P_Z^2(\rho = R_1) + \frac{R_2}{\lambda_S} P_Z^2(\rho = R_2)\right)\right). \tag{4}$$

We introduced longitudinal and lateral extrapolation lengths $\lambda_b \neq \lambda_S$ in Eq.(4). Hereinafter we regard these extrapolation lengths positive.

Variation of the free energy expression $\Delta G = \Delta G_V + \Delta G_S$ yields the following Euler-Lagrange equations with the boundary conditions on the tube faces $z = \pm h/2$ and the sidewalls $\rho = R_{1,2}$ (see e.g. Ref. 6, 22):

$$
\begin{cases}
\alpha_R P_Z + \beta P_Z^3 + \gamma P_Z^5 - \delta\left( \dfrac{\partial^2}{\partial z^2} + \dfrac{1}{\rho}\dfrac{\partial}{\partial \rho}\rho\dfrac{\partial}{\partial \rho} + \dfrac{1}{\rho^2}\dfrac{\partial^2}{\partial \psi^2} \right)P_Z = E_0 + E_Z^d(\rho,\psi,z), \\[2mm]
\left( P_Z + \lambda_b \dfrac{dP_Z}{dz} \right)\Bigg|_{z=h/2} = 0, \qquad \left( P_Z - \lambda_b \dfrac{dP_Z}{dz} \right)\Bigg|_{z=-h/2} = 0, \\[2mm]
\left( P_Z + \lambda_S \dfrac{dP_Z}{d\rho} \right)\Bigg|_{\rho=R_1} = 0, \qquad \left( P_Z - \lambda_S \dfrac{dP_Z}{d\rho} \right)\Bigg|_{\rho=R_2} = 0,
\end{cases}
\tag{5}
$$

The polarization distribution in the ferroelectric phase should be found by direct variational method. This approach (firstly proposed by Glinchuk et. al.[22] for the description of single-domain thin ferroelectric films polar properties) is evolved for ferroelectric nanorods by Morozovska et al.[12] allowing for possible polydomain states appearance in confined particles. Briefly, the domain wall energy is represented by the correlation term $\dfrac{\delta}{2}\left(\nabla P_Z(\rho,z)\right)^2$ in Eq.(1) for the continuous media approximation, polydomain states could be studied with the help of the free energy (1)-(4). However, for their adequate description one should use exact expression for the depolarization field and calculate the polarization distribution in accordance with Eqs.(5) and (B.8-9) self-consistently, namely at $(\lambda_S/R_1) \ll 1$ we obtained:

$$
E_z^d(\rho,\psi,z) = \sum_{n,m=0,s=1}^{\infty} -\frac{4\pi\left(\dfrac{2\pi s}{h}\right)^2 \exp(im\psi)}{\left(\dfrac{2\pi s}{h}\right)^2 + \left(\dfrac{k_{mn}}{R_1}\right)^2}\cos\left(\frac{2\pi s z}{h}\right)\left( J_m\left(\frac{k_{mn}\rho}{R_1}\right) - \frac{J_m(k_{mn})}{N_m(k_{mn})}N_m\left(\frac{k_{mn}\rho}{R_1}\right) \right)P_{mns}^V \quad, \tag{6a}
$$

$$
P_Z(\rho,\psi,z) = \sum_{n,m=0,s=0}^{\infty} \exp(im\psi)\cos\left(\frac{2\pi s z}{h}\right)\left( J_m\left(\frac{k_{mn}\rho}{R_1}\right) - \frac{J_m(k_{mn})}{N_m(k_{mn})}N_m\left(\frac{k_{mn}\rho}{R_1}\right) \right)P_{mns}^V \quad. \tag{6b}
$$

Here $J_m(x)$ and $N_m(x)$ are Bessel and Neiman functions of the $m$-th order respectively; eigenvalues $k_{mn}$ should be found from the lateral boundary conditions

$J_m\left(k_{mn}\dfrac{R_2}{R_1}\right)N_m(k_{mn}) - J_m(k_{mn})N_m\left(k_{mn}\dfrac{R_2}{R_1}\right) = 0$ ; the coefficients $P_{mns}^V$ should be determined.

The inequality $(\lambda_S/R_1) \ll 1$ used in Eq.(6) is valid for typical extrapolation lengths $\lambda_S = 0.3...5\,\text{nm}$ and radiuses $R_1 = 30...500\,\text{nm}$.

Substituting depolarization field (6a) and polarization (6b) into the free energy $\Delta G = \Delta G_V + \Delta G_S$ (see Eqs.(1) and (4)) and integrating over nanoparticle volume, we obtained the free energy with renormalized coefficients for the average polarization, where $P_{mns}^V$ are variational parameters, that should be found from the system of coupled algebraic equations. The cut off the infinite system for $P_{mns}^V$ (i.e. maximal numbers $m$, $n$ and $s$) follows from the free energy minimum that exists allowing for the following reasons. Really, single-domain state posses the highest depolarization field energy, whereas domain splitting leads to its essential decreases. On the other hand, the domain splitting leads to the unlimited increase of the domain wall energy represented by the correlation term $\frac{\delta}{2}(\nabla P_Z(\rho, z))^2$. Thus the optimal number of domains (which can be approximated by harmonics $m$, $n$, $s$ with high accuracy) that corresponds to the free energy minimum exists.

Unfortunately we could not derived analytical expressions for the free energy renormalized coefficients for a polydomain case, only numerical simulations have been performed. However, it is appeared that single-domain state is energetically preferable for infinite tubes and wires, since depolarization field is absent and correlation energy is minimal for single-domain case. In contrast to the finite poly-domain tubes, simple approximate analytical results has been obtained the infinite single-domain ones. Below we report the results.

### III. Phase diagram of the long nanotubes

In Appendix C we derived the interpolation for the paraelectric-ferroelectric transition temperature $T_{CR}(R_1, R_2)$ of the long nanotubes ($h \gg R_1$ so $E_d \to 0$):

$$T_{CR}(R_1, R_2) = T_C + \frac{2Q_{12}}{\alpha_T\left(1 - (R_2/R_1)^2\right)}\left(p_1 + \left(\frac{R_2^2}{R_1^2}\right)p_2\right)\exp\left(-\frac{\Delta R}{R_1 - R_2}\right) - \delta\frac{k_{01}^2(R_1, R_2)}{\alpha_T R_1^2}, \qquad (7)$$

where $k_{01}(R_1, R_2)$ is the first root of the equation $J_0\left(k_{01}\frac{R_2}{R_1}\right)N_0(k_{01}) - J_0(k_{01})N_0\left(k_{01}\frac{R_2}{R_1}\right) = 0$. In fact, the root $k_{01}(R_1, R_2)$ depends on the ratio $R_2/R_1$, in particular $k_{01} \to \infty$ at $R_2/R_1 \to 1$ (see Fig.1C in Appendix C).

The first term in Eq.(7) is the bulk transition temperature, the second term is related to the coupling of radial stress with polarization via electrostriction effect, the third term is caused by correlation effects. The correlation term is always negative and thus only decreases the transition temperature, whereas the electrostriction term in Eq.(7) could be positive or negative depending on the $Q_{12}$ sign. Note, that both signs of $Q_{12}$ are possible for different ferroelectrics, however $Q_{12} < 0$ for most of the perovskite ferroelectrics. Below we demonstrate that increasing of transition temperature and thus ferroelectric properties conservation or even enhancement is possible when $Q_{12} < 0$ and depolarization field is small enough.

Hereinafter we assume that radial pressures $p_1 = -\mu_1/R_1$ and $p_2 = -\mu_2/R_2$ for the sake of specificity (however it is not the only model for them, see e.g. Refs. [24, 26])); put surface tension coefficients equal $\mu_{1,2} = \mu$ for the sake of simplicity. The assumptions essentially simplify Eq.(7), namely:

$$T_{CR}(R_1, R_2) = T_C - \frac{2Q_{12}\mu}{\alpha_T(R_1 - R_2)}\exp\left(-\frac{\Delta R}{R_1 - R_2}\right) - \delta\frac{k_{01}^2(R_1, R_2)}{\alpha_T R_1^2}, \tag{8}$$

Let us make some estimations of the second and third terms in Eq.(8) for perovskites BaTiO$_3$ and PbTiO$_3$. Using parameters $Q_{12} = -0.043\,\mathrm{m^4/C^2}$, $T_C = 400\,\mathrm{K}$ (BaTiO$_3$) and $Q_{12} = -0.046\,\mathrm{m^4/C^2}$, $T_C = 666\,\mathrm{K}$ (PbZr$_{0.5}$Ti$_{0.5}$O$_3$); $\mu_{1,2} = 5 - 50\,\mathrm{N/m}$ (see e.g. Ref. 10) and $\delta = 10^{-19} - 10^{-17}\,\mathrm{m^2}$, we obtained that $\left|\frac{2\mu Q_{12}}{\alpha_T T_C}\right| \approx 2 - 17\,\mathrm{nm}$, $\sqrt{\frac{\delta}{\alpha_T T_C}} \approx 2.3 - 23\,\mathrm{nm}$ for BaTiO$_3$ and $\left|\frac{2\mu Q_{12}}{\alpha_T T_C}\right| \approx 3 - 26\,\mathrm{nm}$, $\sqrt{\frac{\delta}{\alpha_T T_C}} \approx 1.9 - 19\,\mathrm{nm}$ for PbZr$_{0.5}$Ti$_{0.5}$O$_3$ respectively. So both terms are comparable with unity at nanoparticle radius ~ 2-25 nm.

Taking into account that the gradient coefficient $\sqrt{\delta} \sim 0.3...3\,\mathrm{nm}$, i.e. it is of several lattice constants, we introduced the parameters and dimensionless variables that correspond to the lattice constants units:

$$R_\mu = \frac{2\mu Q_{12}}{\alpha_T T_C \sqrt{\delta}}\ , \qquad R_S = \sqrt{\frac{1}{\alpha_T T_C}}, \qquad r_{1,2} = \frac{R_{1,2}}{\sqrt{\delta}}, \qquad \Delta r = \frac{\Delta R}{\sqrt{\delta}}. \tag{9a}$$

In these variables

$$T_{CR}(r_1, r_2) = T_C\left(1 - \frac{R_\mu}{r_1 - r_2}\exp\left(-\frac{\Delta r}{r_1 - r_2}\right) - k_{01}^2\frac{R_S^2}{r_1^2}\right) \tag{9b}$$

Let us underline, that $R_\mu$ sign is determined by the one of electrostriction coefficient $Q_{12}$ (surface tension coefficient $\mu$ is regarded positive). So, the parameter $R_\mu$ is negative for most of perovskites with $Q_{12} < 0$. In accordance with our estimations (see comments to Eq.(7)) we obtained that $R_S \sim 5...10$ and $|R_\mu| \sim 8...80$ depending on the material parameters and surface tension coefficient value respectively.

In Figs.2 we present phase diagrams calculations based on the Eqs.(9). The size effect on the phase diagram for the case when the shape of nanoparticle is fixed ($r_2/r_1 = const$), but its outer radius $r_1$ varies is represented in Figs.2(a, b). Transition temperature $T_{CR}(r_1, r_2)$ vs. the inner radius $r_2$ for different ratios $r_2/r_1$ is represented in Figs.2(c, d).

It is clear from Figs.2(a, b) that transition temperature values and critical radius significantly depend on the nanotube thickness, namely the critical radius is smallest for nanowires ($r_2 = 0$), slightly bigger for "thick" nanotubes ($r_2/r_1 < 0.1$) and biggest the "thin" ones ($r_2/r_1 \approx 1$). The transition temperature $T_{CR}(r_1, r_2)$ tends to the bulk value $T_C$ at $r_1 \to \infty$ for any shape, as it should be expected for the bulk ferroelectric material.

Nanowires and nanotubes reveal noticeable increase of transition temperature ($T_{CR}/T_C > 1$) at $R_\mu < 0$ in the vicinity of the optimal radius $r_0$ (see Figs.2(b), (d)). At $\Delta r << r_1$ the optimal radius $r_0 \approx 2R_S^2/R_\mu$ for nanowires [12] whereas $r_0 \approx 2R_S^2 k_{01}^2 \left(1 - (r_2/r_1)\right)/R_\mu$ for nanotubes (i.e. it depends on the ratio $r_2/r_1$ only, since $k_{01}^2(r_1, r_2) \equiv k_{01}^2(r_2/r_1)$). For thin nanotubes ($r_2/r_1 \approx 1$) one obtains that $k_{01}^2 \sim \left(1 - (r_2/r_1)\right)^{-2}$, so maximum corresponds to the radius $r_0 \sim 2R_S^2/R_\mu \left(1 - (r_2/r_1)\right)$. The enhancement of transition temperature is caused by the competition between the radial stress that increases a paraelectric-ferroelectric transition temperature via negative electrostriction and the correlation effect that decreases the transition temperature. Note, that the polarization enhancement in thin nanotubes could be explained by peculiarities of stress size dependence given by Eq.(2b), namely for nanotubes it reaches the maximal value $\sim \mu/\Delta r$ at $r_1 - r_2 = \Delta r$, not disappears proportionally to $1/r_1$ as valid for nanowires.

No enhancement was obtained for $R_\mu > 0$ ($T_{CR}/T_C$ is always smaller than unity and monotonically increases with outer radius increase (see Fig.2(a)) and decreases with inner radius increase (see Fig.2(c)). For $R_\mu > 0$ the transition temperature $T_{CR}(r_1, r_2)$ is the highest for the tube with the biggest outer radius and smallest for the thinnest one. Really under the condition $R_\mu > 0$ both the correlation effect and radial stress decrease a paraelectric-ferroelectric transition temperature.

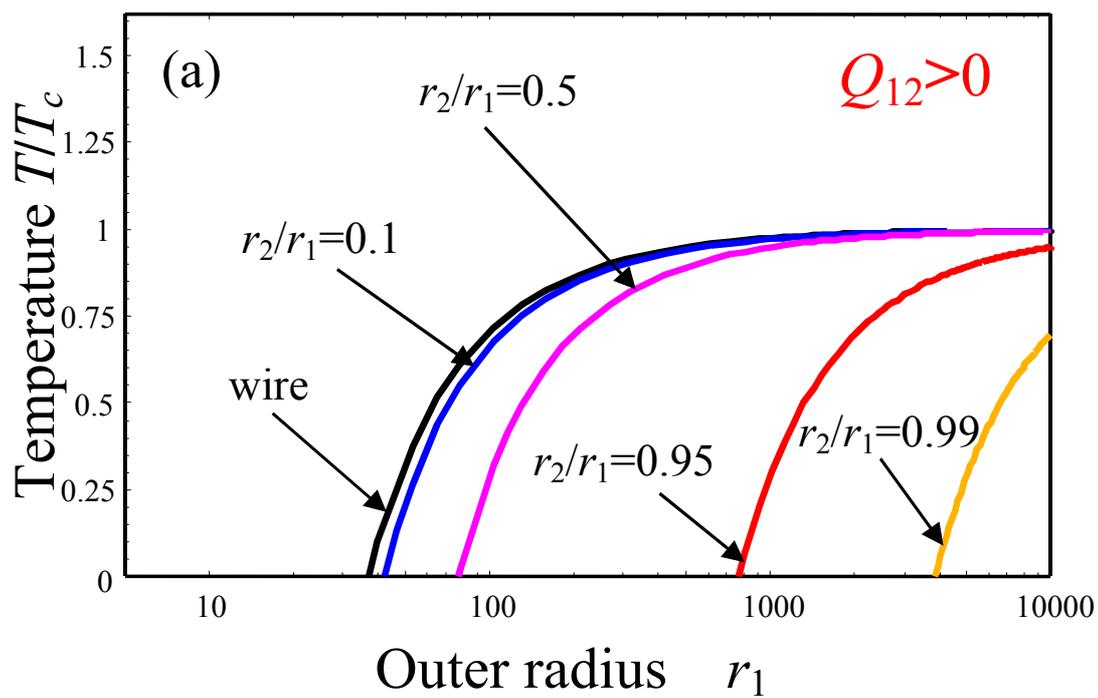

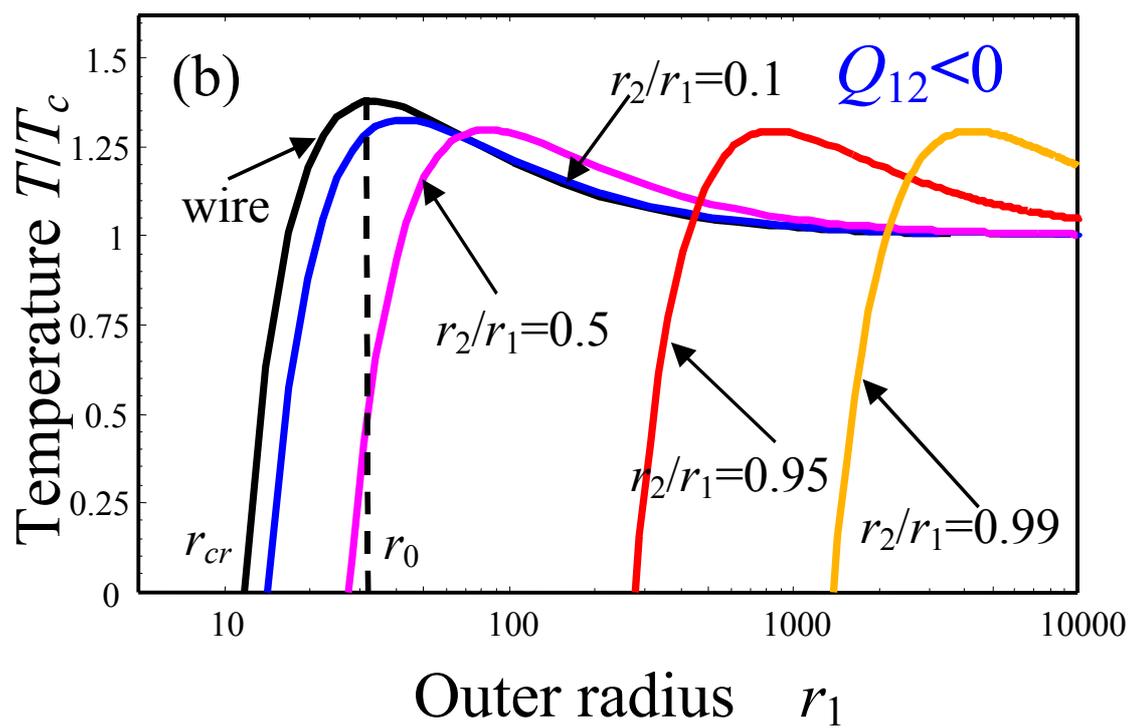

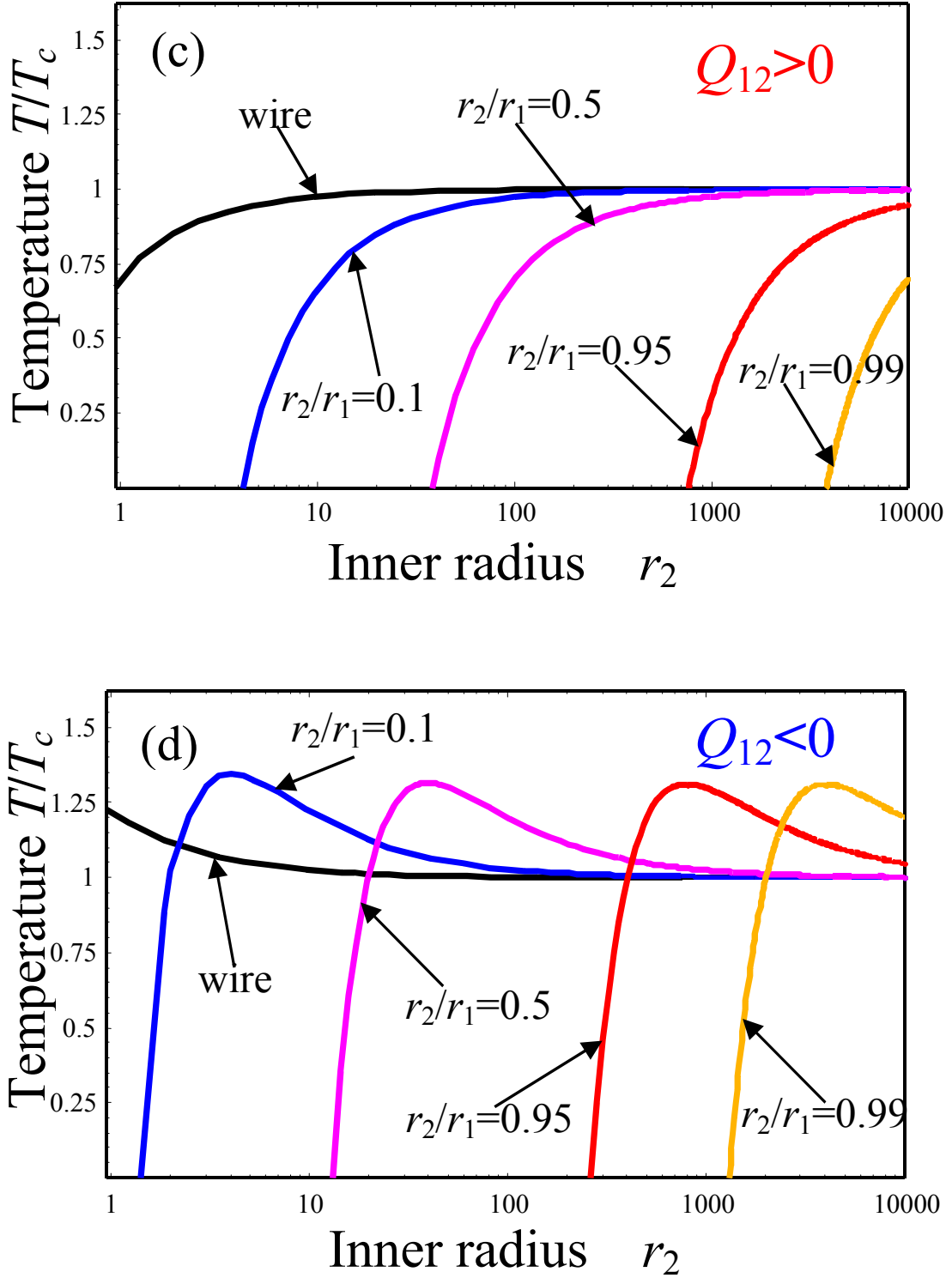

FIG.2 (Color online) Transition temperature $T_{CR}(r_1, r_2)$ vs. outer radius $r_1$ (a,b) and inner radius $r_2$ (c,d) for different ratios $r_2/r_1$: <0.01 (wire); 0.1; 0.5; 0.95; 0.99 (figures near the curves). Other parameters: $\alpha_T = 2.95 \cdot 10^{-5}$, $T_C = 666$ K, $R_S \approx 7$, $\Delta r = 5$ and $R_\mu = \pm 25$ correspond to PbZr$_{0.5}$Ti$_{0.5}$O$_3$.

## IV. Polar properties of the long nanotubes and nanowires

The novel results have been obtained in the case $R_\mu < 0$: long nanotubes reveal noticeable increase of transition temperature (see Figs.2 (b,d)). In this section we demonstrate, that under the condition $R_\mu < 0$ long nanotubes also posses enhanced polar properties, namely they have higher spontaneous polarization and piezoelectric coefficient in comparison with a bulk sample.

For long enough nanotubes we derived approximate analytical expressions for the free energy with renormalized coefficients, namely:

$$\Delta G \approx \frac{\alpha_T}{2}(T - T_{CR}(R_1, R_2))P_n^2 + \frac{\beta}{4}P_n^4 + \frac{\gamma}{6}P_n^6 - P_n E_0 . \qquad (10)$$

Here $T_{CR}(R_1, R_2)$ is given by Eq.(7).

The free energy (10) has conventional form of power series on the averaged polarization. From the free energy one immediately obtains the average spontaneous polarization, coercive field, ferroelectric and dielectric hysteresis loops as well as piezoelectric coefficient after solving algebraic equations. Namely, for the ferroelectrics with the second order phase transition: spontaneous polarization $P_{nS} = \sqrt{-\alpha_T(T - T_{CR}(R_1, R_2))/\beta}$ and thermodynamic coercive field

$E_C^n = \frac{2P_{nS}}{3\sqrt{3}}\alpha_T(T - T_{CR}(R_1, R_2))$. Quasi-equilibrium ferroelectric and dielectric hysteresis loops have

been calculated from the Landau-Khalatnikov equation [16]:

$$-\Gamma\frac{\partial P_n}{\partial t} = \alpha_T(T - T_{CR}(R_1, R_2))P_n + \beta P_n^3 + \gamma P_n^5 - E_{0m}\sin(\omega t), \qquad (11a)$$

$$-\Gamma\frac{\partial \chi_{33}}{\partial t} = \alpha_T(T - T_{CR}(R_1, R_2))\chi_{33} + 3\beta P_n^2\chi_{33} + 5\gamma P_n^4\chi_{33} - 1 . \qquad (11b)$$

Where $P_n$ is the nanoparticle polarization and $\chi_{33} = \frac{\partial P_n}{\partial E_0}$ is its dielectric susceptibility,

$E_0 = E_{0m}\sin(\omega t)$ is the quasi-static applied electric field; external field period $2\pi/\omega$ is regarded small in comparison with inner relaxation time $\Gamma/(\alpha_T T_C)$ .

The piezoelectric tensor coefficients $d_{kij}$ are proportional to the polarization and susceptibility values $d_{kij}(\mathbf{r}) \sim Q_{ijml}\chi_{lk}(\mathbf{r})P_m(\mathbf{r})$,[30] i.e. $d_{33} = 2Q_{11}\chi_{33}P_3$ , $d_{31} = 2Q_{12}\chi_{33}P_3$ and $d_{15} = 2Q_{44}\chi_{11}P_3$ in the case, when only the component $P_3 \neq 0$ .

The dependence of spontaneous polarization $P_{nS}$ on the temperature $T$ is depicted in Figs.3 for nanowires and nanotubes of different outer radius $r_1$ and thickness determined by the ratio $r_2/r_1$ .

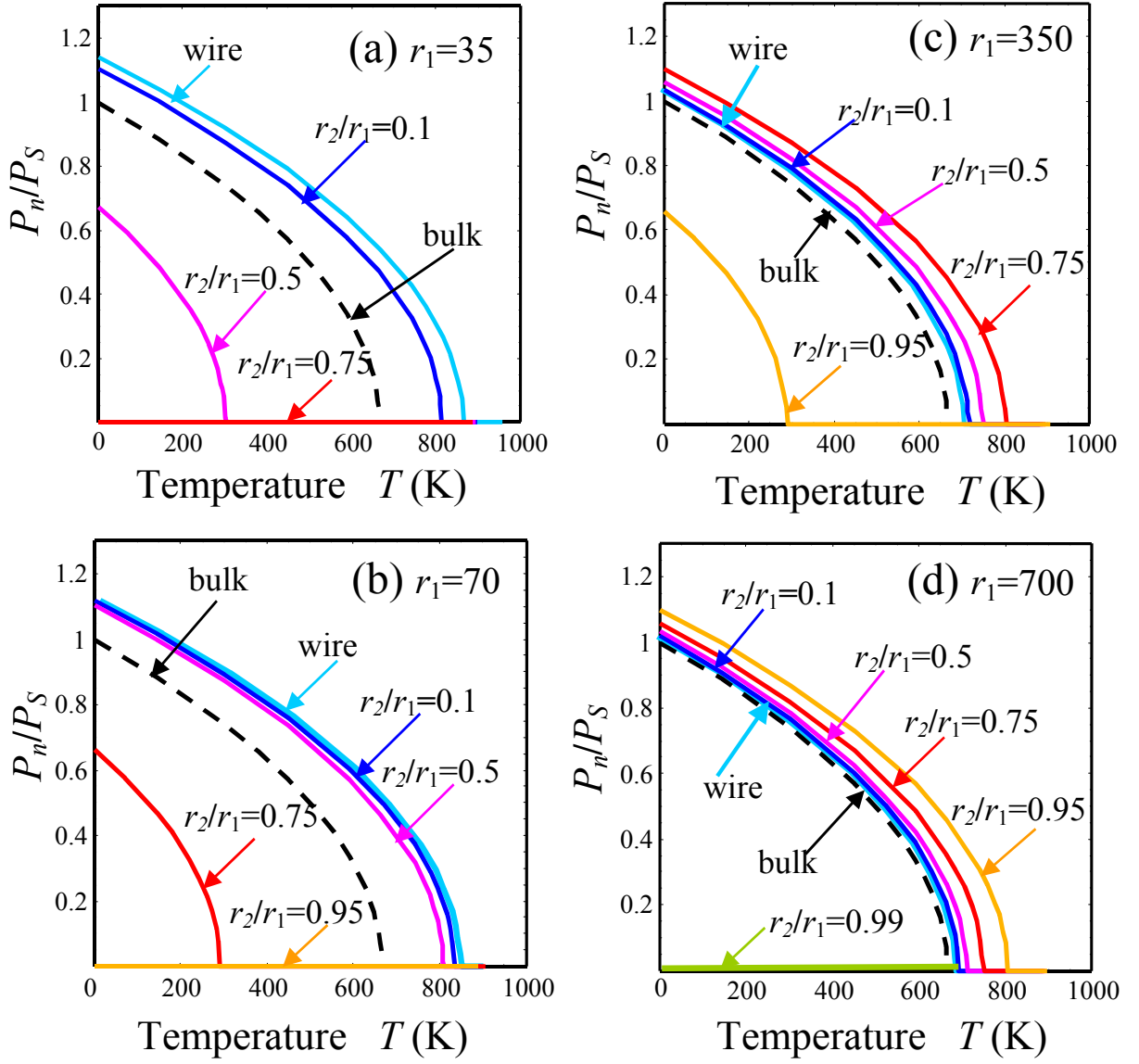

FIG.3 (Color online) Spontaneous polarization $P_n(T)/P_S(0)$ vs. temperature for different ratios $r_2/r_1$ (figures near the curves) and outer radius $r_1 = 35$ (a), $r_1 = 70$ (b), $r_1 = 350$ (c), $r_1 = 700$ (d). Other parameters: $\alpha_T = 2.95 \cdot 10^{-5}$, $T_C = 666$ K, $R_S \approx 7$, $\Delta r = 5$ and $R_\mu = -25$ correspond to PbZr$_{0.5}$Ti$_{0.5}$O$_3$.

It is clear that spontaneous polarization value is higher that the bulk one for nanowires and nanotubes of definite thicknesses determined by the ratio $r_2/r_1$. Namely, for small outer radius ($r_1 < 50$) nanowires and thick nanotubes ($r_2/r_1 < 0.1$) posses enhanced spontaneous polarization ($P_n/P_S > 1$) existing the wider temperature range (e.g. $0 \le T \le 1.3 T_C$) in comparison with a bulk sample, whereas thin nanotubes ($r_2/r_1 > 0.5$) reveal depressed spontaneous polarization ($P_n/P_S < 1$) existing in more narrow temperature range (e.g. $0 \le T \le 0.5 T_C$) in comparison with a bulk (see Fig.3(a)). For chosen material parameters and $r_1 = 35$ ferroelectricity disappears at $r_2/r_1 > 0.65$. For big nanoparticle sizes ($r_1 > 500$) both nanowires and nanotubes of different thickness posses slightly enhanced spontaneous polarization in comparison with a bulk sample, however thin nanotubes ($0.5 < r_2/r_1 < 0.95$) reveal the highest polarization and transition temperature in comparison with nanowires and thin tubes (see Fig.3(d)). For big ultrathin tubes ($r_1 > 500$) ferroelectricity disappears only at $r_2/r_1 > 0.99$, when the positive correlation term becomes too high and stress relaxation appears (see comments to Eq.(2b)). Note, that the curves for different ratios $r_2/r_1$ tend to the bulk one and change their order with outer radius increase (compare plots (a)-(d)). The polarization enhancement in thin nanotubes of big outer radius could be explained by peculiarities of stress size dependence given by Eq.(2b), namely for nanotubes it reaches the maximal value $\sim \mu/\Delta r$ at $r_1 - r_2 = \Delta r$.

The dependences of spontaneous polarization $P_{nS}$ and thermodynamic coercive field $E_C^n$ on the nanotube outer radius are depicted in Figs.4. It is clear from the figure that the regions with spontaneous polarization $P_{nS}$ higher than its bulk value $P_S(T)$ always exist at $R_\mu < 0$ (see Figs.4(a,c,e)). In the same region or radiuses the coercive field $E_C^n$ is higher than the bulk value $E_C(T)$ (see Figs.4(b,d,f)). Moreover, the closest is the temperature $T$ to the transition value $T_C$ in the bulk, the higher are the ratios $P_{nS}(T)/P_S(T)$ and $E_C^n(T)/E_C(T)$. It is clear from the Figs.4(c-f), that at room temperature $T = 300\,\mathrm{K}$, radius $r_1 = 20$ and $r_2/r_1 = 0.01$ the ratios $P_n/P_S \approx 1.3$ and $E_C^n/E_C \approx 2.3$, whereas $P_n/P_S \approx 2.5$ and $E_C^n/E_C \approx 12$ at $T = 600\,\mathrm{K}$. These effects are the most pronounced for nanowires ($r_2/r_1 \to 0$) and less pronounced for thin tubes ($r_2/r_1 \to 1$). It is interesting, that the coercive field firstly increases with the tube outer radius increase, quickly reaches the maximum and then decreases tending to the bulk value with the tube outer radius increase.

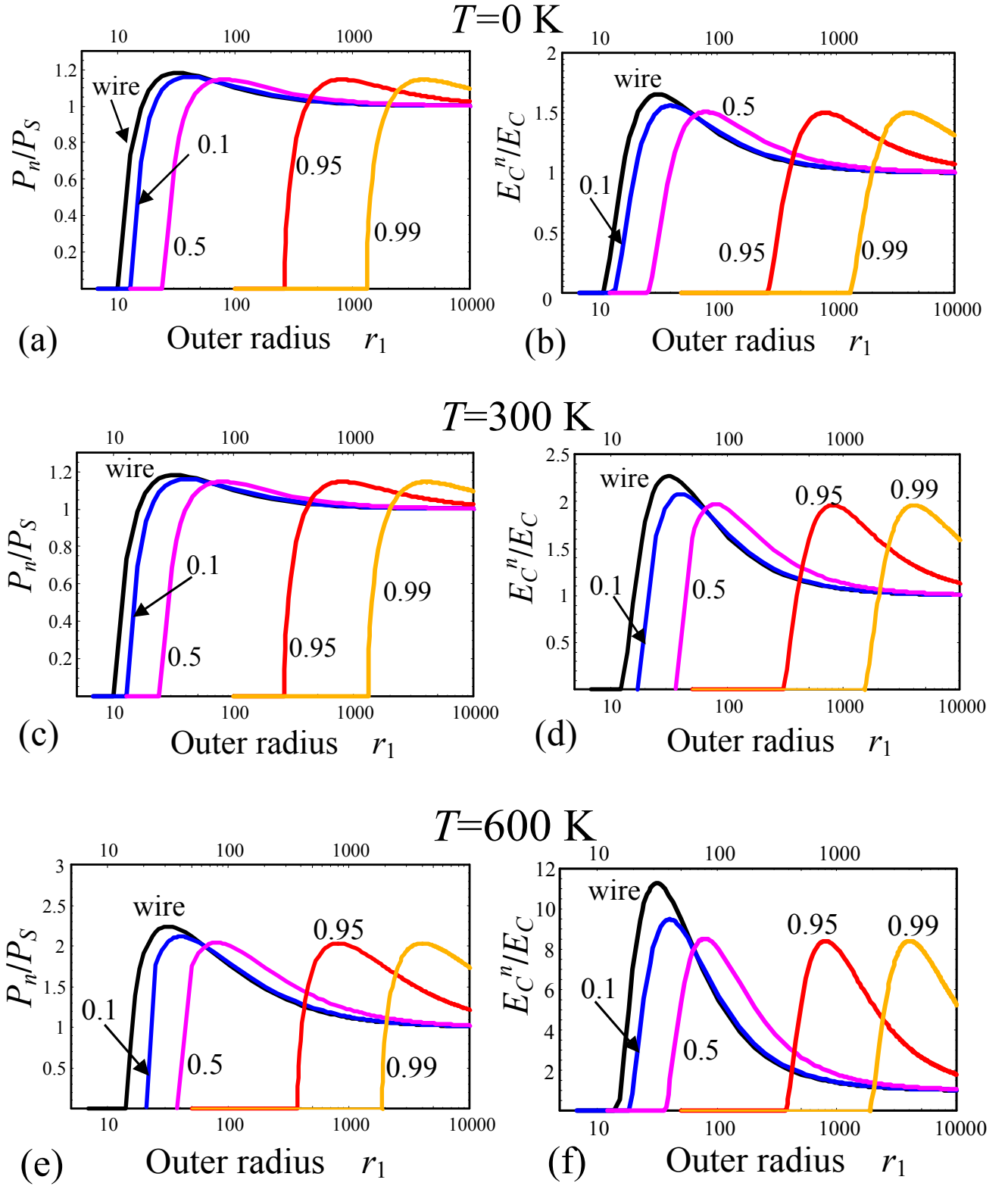

FIG.4 (Color online) Spontaneous polarization $P_{nS}(T)/P_S(T)$ (a, c, e) and thermodynamic coercive field $E_C^n(T)/E_C(T)$ vs. outer radius $r_1$ (b, c, d) for different ratios $r_2/r_1$ : <0.01 (wire); 0.1; 0.5; 0.95; 0.99 (figures near the curves); and temperatures $T = 0$ K (a, b), $T = 300$ K (c, d), $T = 600$ K (e, f). Other parameters: $\alpha_T = 2.95 \cdot 10^{-5}$, $T_C = 666$ K, $R_S \approx 7$, $\Delta r = 5$ and $R_\mu = -25$ correspond to PbZr$_{0.5}$Ti$_{0.5}$O$_3$.

Thermodynamic ferroelectric hysteresis loops are shown in Figs. 5.

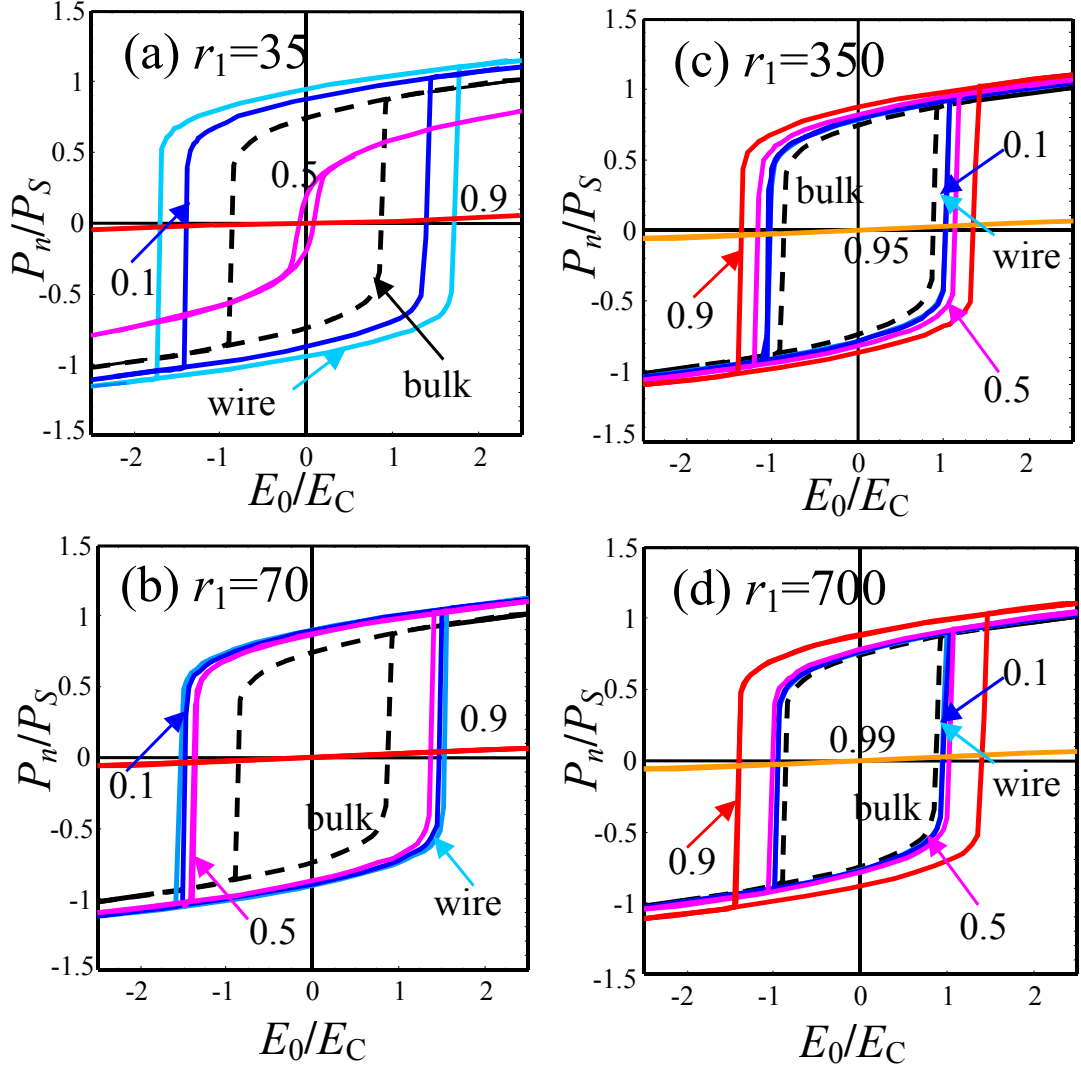

FIG.5 (Color online) Thermodynamic ferroelectric hysteresis loops ($P_n/P_S$ ($T=0$) vs. $E_0/E_C$ ($T=0$)) of nanotubes with different outer radius $r_1=35$ (a), $r_1=70$ (b), $r_1=350$ (c), $r_1=700$ (d) and ratios $r_2/r_1$ (figures near the curves). Other parameters: $T=300$ K, $T_C=666$ K, $\alpha_T=2.95 \cdot 10^{-5}$, $R_S \approx 7$, $\Delta r=5$ and $R_\mu=-25$ correspond to $PbZr_{0.5}Ti_{0.5}O_3$.

At room temperature ferroelectric hysteresis loops of the small-sized ($r_1<50$) nanowires and thick nanotubes ($r_2/r_1<0.1$) be could be noticeably wider and higher that the one for a bulk sample, whereas the loop disappears for thick tubes ($r_2/r_1>0.5$) (see Fig.5(a)). For large-sized nanoparticles ($r_1>500$) both nanowires and nanotubes of different thickness posses slightly enlarged hysteresis loops in comparison with a bulk sample, however thin nanotubes ($0.5<r_2/r_1<0.9$) reveal the highest remnant polarization and thermodynamic coercive field in comparison with nanowires and thin tubes (see Fig.5(d)). For large ultrathin tubes ($r_1>500$) ferroelectric loops disappear only at $r_2/r_1>0.99$, when the positive correlation term becomes too high and stress relaxation appears. Similarly to the situation depicted in Figs.3, the curves for different ratios $r_2/r_1$ tend to the bulk one and change their order with outer radius increase (compare plots (a)-(d)). Both remnant polarization and coercive field increase in thin nanotubes of big outer radius could be explained by the peculiarities of stress size dependence given by Eq.(2b).

Thermodynamic piezoelectric hysteresis loops for $d_{33}$ are presented in Figs.6. Sharp maximums near the coercive field originated from the dielectric permittivity maxima (see insets in Figs.6).

Similarly to ferroelectric hysteresis, piezoelectric loops of the small-sized ($r_1 = 35$) nanowires and thick nanotubes ($r_2/r_1 < 0.1$) could be noticeably wider and higher that the one for a bulk sample, whereas the loop smears and disappears for thick tubes ($r_2/r_1 > 0.5$) at room temperature (compare Fig.6(a) with Fig.5(a)). For large-sized nanoparticles ($r_1 = 700$) both nanowires and nanotubes of different thickness posses enlarged hysteresis loops in comparison with a bulk sample, however thin nanotubes ($0.5 < r_2/r_1 < 0.9$) reveal the highest thermodynamic coercive field in comparison with nanowires and thin tubes (see Fig.6(d)). For big ultrathin tubes ($r_1 > 500$) ferroelectric loops disappear only at $r_2/r_1 > 0.99$. Similarly to the situation depicted in Figs.3, the curves for different ratios $r_2/r_1$ tend to the bulk one and change their order with outer radius increase (compare plots (a)-(d)). Coercive field increase in thin nanotubes of big outer radius could be explained by peculiarities of stress size dependence given by Eq.(2b).

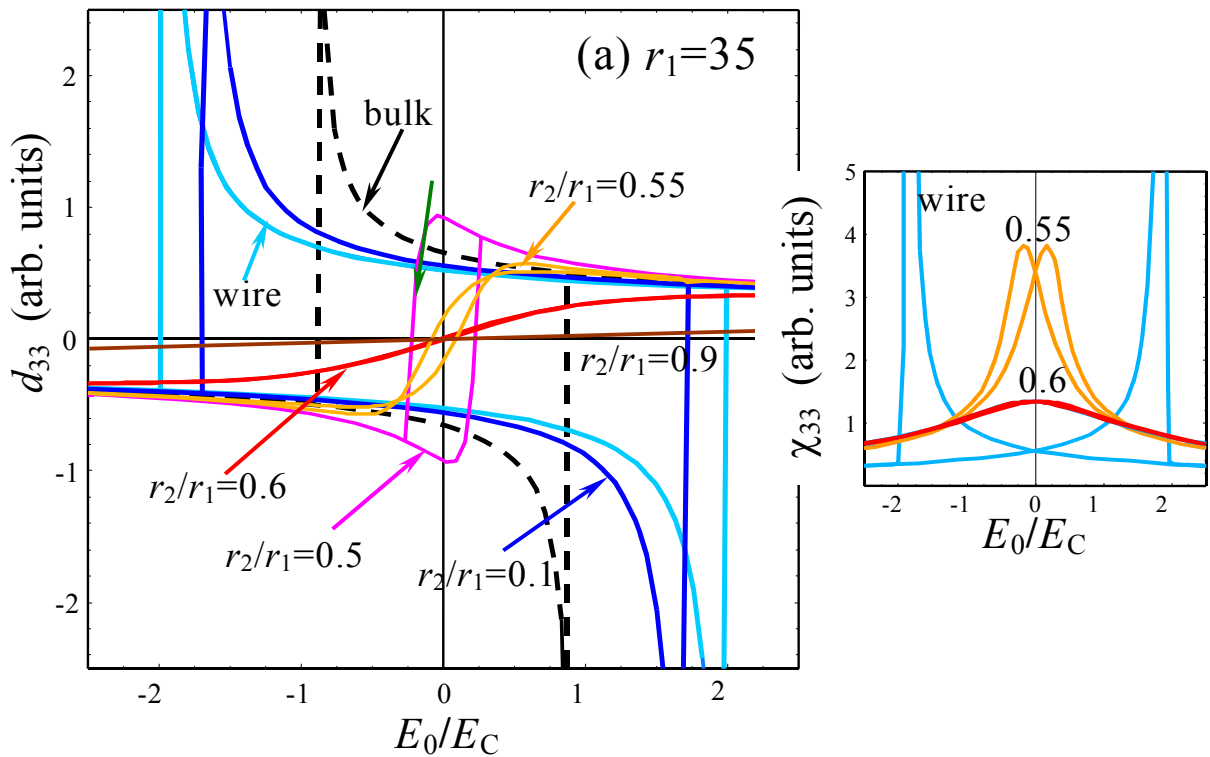

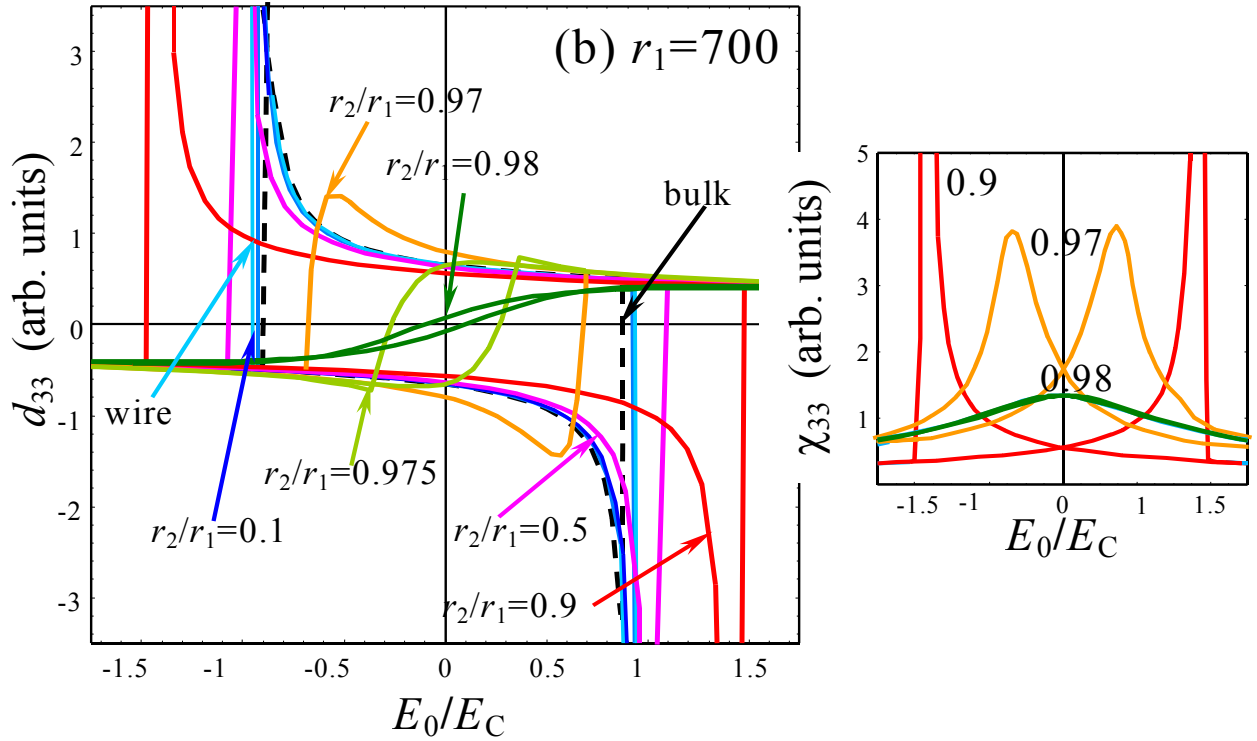

FIG.6 (Color online) Quasi-static piezoelectric hysteresis loops $d_{33}$ vs. $E_0/E_C(T=0)$ for "small tubes" with outer radius $r_1 = 35$ (a) and "large tubes" with outer radius $r_1 = 700$ (b). Figures near the curves denote ratios $r_2/r_1$. Dielectric susceptibility hysteresis is shown in the inset. Other parameters: $\alpha_T = 2.95 \cdot 10^{-5}$, $T_C = 666$ K, $T = 300$ K, $R_S \approx 7$, $\Delta r = 5$ and $R_\mu = -25$ correspond to PbZr$_{0.5}$Ti$_{0.5}$O$_3$.

## V. Comparison with experiment and discussion

Polar properties enhancement in confined RS nanorods was reported by Yadlovker and Berger [1] and partly explained earlier.[12, 13] Below additional comparison of ferroelectric hysteresis loop obtained in RS by Yadlovker et. al[1] with our simulations is shown in Fig.7.

Let us notice, that Yadlovker and Berger[1] observed ferroelectric domains with walls oriented along the finite RS nanorod polar axis. For these states adequate description one should use exact expression (6a) for the depolarization field and calculate the polarization distribution (6b) using direct variational method self-consistently as described earlier. Our numerical simulations proved that hysteresis loops (both ferroelectric and piezoelectric) become more smeared; in particular sharp maximums near thermodynamic coercive field in piezoelectric hysteresis loops disappear, since domain splitting leads to the noticeable smearing of dielectric permittivity maximums.

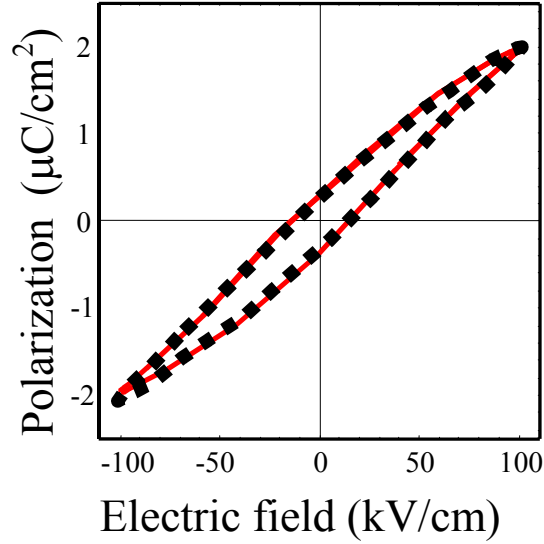

FIG.7 (Color online) Ferroelectric hysteresis loop in RS nanorods of radius 30nm. Squares are experimental data obtained of Yadlovker and Berger[1] at applied field frequency 30kHz, solid curve is our fitting at dimensionless frequency $\omega\Gamma/(\alpha_T T_C) = 0.07$, $T_C = 297\,\mathrm{K}$, $T = 300\,\mathrm{K}$, $R_1 \approx 10$ (i.e. $\sqrt{\delta} = 3$ nm), $R_S \approx 7.4$, $\Delta r = 5$ and $R_\mu = -0.5$ that corresponds to RS.

Recently Morrison et. al.[4],[5] demonstrated that long Pb(Zr,Ti)O$_3$ and BaTiO$_3$ nanotubes posses perfect piezoelectric properties. Let us remind, that the nanoparticle surface displacement components $u_i$ caused by inhomogeneous electric field of a charged AFM probe are registered in Piezoresponse Force Microscopy (PFM, see e.g. Ref. 31). Appeared that measured effective piezoelectric response value $d_{33}^{eff} = u_3/U$ ($u_3$ is a, $U$ is the voltage applied to the AFM probe) is close or higher than the bulk ones. Hereinafter we assume that effective piezoelectric response of the particle $d_{33}^{eff}$ is proportional to the piezoelectric tensor coefficients $d_{ij}$ convoluted with definite elastic Green function. The piezoresponse of the charged AFM tip that touched the surface in the center of the uniformly polarized cylindrical domains is considered in Ref. 32. Extending the results for a tube, as a superposition of two coaxial oppositely polarized nested cylindrical domains, the authors obtained that

$$d_{33}^{eff} = t_{13}(R_1, R_2, \gamma)d_{31} + t_{51}(R_1, R_2, \gamma)d_{15} + t_{33}(R_1, R_2, \gamma)d_{33} \ . \qquad (12)$$

Where rather cumbersome functions $t_{13}(R_1, R_2, \gamma)$ depend only on tube radiuses, dielectric anisotropy coefficient $\gamma = \sqrt{\varepsilon_{33}/\varepsilon_{11}}$ and probe electric field distribution.[32] Thus, the effective piezoresponse $d_{33}^{eff}$ polarization dependence is fully determined by piezoelectric coefficients $d_{33} = 2Q_{11}\chi_{33}P_3$, $d_{31} = 2Q_{12}\chi_{33}P_3$ and $d_{15} = 2Q_{44}\chi_{11}P_3$ in the case, when only the component $P_3 \neq 0$.

However, Eq.(12) was derived in a rigid model for polarization $P_Z \equiv P_S$, so the relation $d_{ij}^{eff} \sim \chi_{kj}P_Z(\rho, \psi)$ is not rigorous for the definite distributions of polarization $P_Z(\rho, \psi)$ and susceptibility $\chi_{33}(\rho, \psi)$. Unfortunately we could not obtain simple analytical expression for $d_{33}^{eff}$ allowing for polarization spatial distribution (by the way Eqs.(11) is obtained in homogeneous external field $E_0$, not in the inhomogeneous one caused by charged probe) and therefore it is questionable to compare simulated piezoelectric coefficient loops with effective piezoresponse ones exactly. However, it is obvious, that within the framework of linear elasticity theory effective piezoresponse $d_{33}^{eff}$ is proportional to the nanoparticle average polarization $P_n$ and susceptibilities $\chi_{ij}$ as following.

$$d_{33}^{eff}(U) \sim P_n(U)\left(\chi_{33}(U) + \vartheta\right) \tag{13}$$

Hereinafter $U \sim E_0 l$ is applied voltage and $\vartheta \sim d_{51}\chi_{11}/d_{33}$ is a fitting parameter.

Despite aforementioned warnings we dared to compare the piezoresponse loop shape obtained for PbZr$_{52}$Ti$_{48}$O$_3$ nanotube[5] and BaTiO$_3$ honeycomb[15] with our modelling in Figs.8-9.

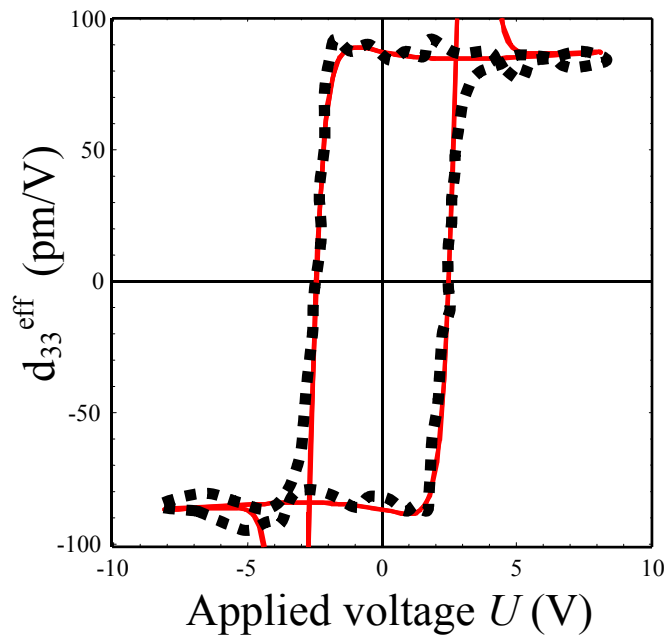

FIG.8 (Color online) Effective piezoresponse $d_{33}^{eff}$ of PbZr$_{52}$Ti$_{48}$O$_3$ nanotube (outer diameter 700nm, wall thickness 90nm, length about 30μm) vs. applied voltage $U$; the loop was centered. Squares are experimental data of Morrison et al.,[5] solid curve is our fitting (13) for $R_1 \approx 700$, $R_2 \approx 610$ (i.e. $\sqrt{\delta} = 1\,\text{nm}$), $\alpha_T = 2.95 \cdot 10^{-5}$, $T_C = 666\,\text{K}$, $T = 300\,\text{K}$, $R_S \approx 7$, $\Delta r = 5$, $R_\mu = -5$, $\vartheta = 0.25$ and dimensionless frequency $\omega\Gamma/\left(\alpha_T T_C\right) = 0.15$. Also we used bulk values $d_{31} = -93.5\,\text{pm/V}$, $d_{15} = 494\,\text{pm/V}$, $d_{33} = 220\,\text{pm/V}$, $\varepsilon_{11} = 1180$, $\varepsilon_{33} = 730$ [14] ($\gamma \approx 0.79$).

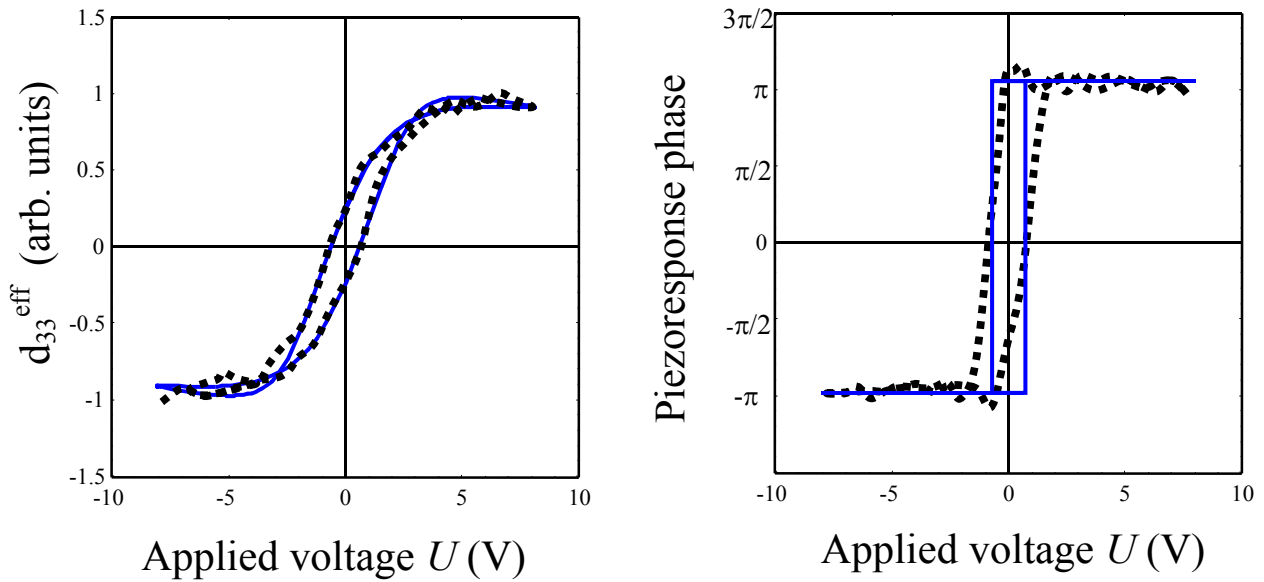

FIG.9 (Color online) Effective piezoresponse $d_{33}^{eff}$ of nanotube-patterned BaTiO$_3$ "honeycomb" (inner radius 50-100 nm, film thickness about 200-300nm) vs. applied voltage $U$; the loop was centered. Squares are experimental data of Poyato et al.,[15] solid curve is our fitting (13) at $R_2 = 50\,\text{nm}$, $R_1 = 62\,\text{nm}$, $\alpha_T = 5.69 \cdot 10^{-5}$, $T_C = 400\,\text{K}$, $T = 300\,\text{K}$, $R_S \approx 6$, $\Delta r = 5$, $R_\mu = -5$, $\vartheta$ is negligibly small and dimensionless frequency $\omega\Gamma/(\alpha_T T_C) = 0.15$. Also we used bulk values $d_{31} = -34.5\,\text{pm/V}$, $d_{33} = 86\,\text{pm/V}$, $d_{15} = 392\,\text{pm/V}$, $\varepsilon_{11} = 2920$, $\varepsilon_{33} = 168$,[14] ($\gamma \approx 0.24$).

Note, that Poyato et. al[15] reported that nanotube-patterned film thickness was about 200-300 nm, thus the honeycomb could be poly-domain. The domain walls existence was reported by Nagarajan et al.[33] Domain wall motion may lead to additional piezoresponse loop smearing in finite nanoparticles.

Let us remind, that RS is known to be improper ferroelastic – ferroelectric, therefore our consideration can be applied to the material only in the temperature range where RS ferroelectric properties can be described by the phenomenological expansion (1) over polarization powers. Moreover, we neglected the piezoelectric effect with respect to the shear stress in the paraelectric phase of RS since the radial pressure creates no tangential stresses.

Also semiconductor properties of BaTiO$_3$ and Pb(Zr,Ti)O$_3$ as well as the interfacial phenomena between the ferroelectric and porous [34] could influence on the loop sizes and shape, in particular the latter causes loop shift and/or imprint.

Despite aforementioned remarks and mute points our fitting is in a surprisingly good agreement with observed ferroelectric and local piezoresponse hysteresis loops. However, in order to avoid uncertainty the radial stress influence on the ferroelectricity conservation in confined nanotubes should be verified experimentally.[*]

Really, the compressive radial stress determined by the term (2b) is proportional to $Q_{12}f(R_1, R_2)$ ($Q_{12}$ is regarded known). In order to study the surface pressure influence one has to vary this term experimentally. So, one could tune the surface forces by changing the nanotube radii $R_{1,2}$, e.g. by fitting the average porous sizes, sieves, precursor concentration, drying conditions *etc*. The quantitative comparison of the measured dependence $T_{CR}(R_1, R_2)$ with the one calculated from Eqs.(7)-(8) would be extremely desirable in order to verify our model. If it is appeared that Eqs.(9b) is valid at the reasonable values of fitting parameters (9a), one could say that the radial stress as well as depolarization field decrease in long cylindrical nanoparticles are the keys to the ferroelectricity conservation and enhancement in nanotubes and nanowires.

## Appendix A

The free energy expansion on polarization $\mathbf{P} = (0, 0, P_3)$ and stress $\sigma_i$ powers has the form[17, 24]:

$$F = a_1 P_3^2 + a_{11} P_3^4 + a_{111} P_3^6 - Q_{11}\sigma_3 P_3^2 - Q_{12}(\sigma_1 + \sigma_2)P_3^2 - $$
$$- \frac{1}{2}s_{11}(\sigma_1^2 + \sigma_2^2 + \sigma_3^2) - s_{12}(\sigma_1\sigma_2 + \sigma_1\sigma_3 + \sigma_3\sigma_2) - \frac{1}{2}s_{44}(\sigma_4^2 + \sigma_5^2 + \sigma_6^2) \quad \text{(A.1)}$$

Hereinafter we use Voigt notation $\sigma_i$ or matrix notation $\sigma_{nm}$ (xx=1, yy=2, zz=3, zy=4, zx=5, xy=6) when it necessary.

Firstly let us calculate the $\sigma_i$ components caused by the uniform radial pressure related to the effective surface pressures $p_1$ and $p_2$.[29, 10] This Lame's problem is discussed in details elsewhere.[35]

The conditions of mechanical equilibrium $n_i\sigma_{ij} = +p_j$ on the surface of cylindrical solid body in the cylindrical coordinates $(\rho, \psi, z)$ have the following form:

$$\sigma_{\rho\rho}\big|_{\rho=R_1} = p_{1\rho}, \quad -\sigma_{\rho\rho}\big|_{\rho=R_2} = p_{2\rho}, \quad \sigma_{\rho\psi}\big|_{\rho=R_{1,2}} = 0, \quad \sigma_{\rho z}\big|_{\rho=R_{1,2}} = 0,$$
$$\sigma_{zz}\big|_{z=\pm h/2} = 0, \quad \sigma_{z\rho}\big|_{z=\pm h/2} = 0, \quad \sigma_{z\psi}\big|_{z=\pm h/2} = 0 \quad \text{(A.2)}$$

The conditions of mechanical equilibrium $\partial\sigma_{ij}/\partial x_i = 0$ in the bulk of solid body are the following:

---

[*] Also [34] supposes, that the polarization enhancement observed by the researchers in Ref. 1 is a stress effect caused by clamping of the nanoparticles by the porous matrix rather than a mere size effect exists. Really, the phase transition temperature increase takes place in the epitaxial films of perovskite ferroelectrics due to the misfit strain between film and its substrate.[17] To our mind, lateral clamping of the nanoparticles by the porous matrix should contribute into the effective surface stress, thus actually we deal with mixed "stress-size" effect.

$$\begin{cases} \dfrac{\partial \sigma_{zz}}{\partial z} + \dfrac{\partial \sigma_{z\rho}}{\partial \rho} + \dfrac{\sigma_{z\rho}}{\rho} + \dfrac{1}{\rho}\dfrac{\partial \sigma_{z\psi}}{\partial \psi} = 0, \\[3mm] \dfrac{\partial \sigma_{\rho\rho}}{\partial \rho} + \dfrac{\sigma_{\rho\rho} - \sigma_{\psi\psi}}{\rho} + \dfrac{1}{\rho}\dfrac{\partial \sigma_{\rho\psi}}{\partial \psi} + \dfrac{\partial \sigma_{z\rho}}{\partial z} = 0, \\[3mm] \dfrac{1}{\rho}\dfrac{\partial \sigma_{\psi\psi}}{\partial \psi} + \dfrac{\partial \sigma_{z\psi}}{\partial z} + \dfrac{\partial \sigma_{\rho\psi}}{\partial \rho} + 2\dfrac{\sigma_{\rho\psi}}{\rho} = 0. \end{cases} \tag{A.3}$$

It is seen that boundary and equilibrium conditions (A.2) and (A.3) can be fulfilled with

$$\sigma_{\rho\rho} = \frac{1}{1 - (R_2/R_1)^2}\left( p_{1\rho}\left(1 - \frac{R_2^2}{\rho^2}\right) + \frac{R_2^2}{R_1^2}\left(1 - \frac{R_1^2}{\rho^2}\right)p_{2\rho}\right),$$

$$\sigma_{\psi\psi} = \frac{1}{1 - (R_2/R_1)^2}\left( p_{1\rho}\left(1 + \frac{R_2^2}{\rho^2}\right) + \frac{R_2^2}{R_1^2}\left(1 + \frac{R_1^2}{\rho^2}\right)p_{2\rho}\right), \tag{A.4a}$$

$$\sigma_{\rho\psi} = 0, \quad \sigma_{\rho z} = 0, \quad \sigma_{zz} = 0, \quad \sigma_{z\psi} = 0 \tag{A.4b}$$

The tensor components in Cartesian coordinates can be found from relations:

$$\begin{pmatrix} \sigma_{xx} & \sigma_{xy} & \sigma_{xz} \\ \sigma_{yx} & \sigma_{yy} & \sigma_{yz} \\ \sigma_{zx} & \sigma_{zy} & \sigma_{zz} \end{pmatrix} = \begin{pmatrix} \cos(\psi) & -\sin(\psi) & 0 \\ \sin(\psi) & \cos(\psi) & 0 \\ 0 & 0 & 1 \end{pmatrix} \cdot \begin{pmatrix} \sigma_{\rho\rho} & \sigma_{\rho\psi} & \sigma_{\rho z} \\ \sigma_{\psi\rho} & \sigma_{\psi\psi} & \sigma_{\psi z} \\ \sigma_{z\rho} & \sigma_{z\psi} & \sigma_{zz} \end{pmatrix} \cdot \begin{pmatrix} \cos(\psi) & \sin(\psi) & 0 \\ -\sin(\psi) & \cos(\psi) & 0 \\ 0 & 0 & 1 \end{pmatrix} \tag{A.5}$$

Allowing for Eq. (A.4), expression (A.5) leads to

$$\sigma_{xx} = \cos^2(\psi)\sigma_{\rho\rho} + \sin^2(\psi)\sigma_{\psi\psi} \neq 0,$$

$$\sigma_{yy} = \sin^2(\psi)\sigma_{\rho\rho} + \cos^2(\psi)\sigma_{\psi\psi} \neq 0,$$

$$\sigma_{xy} = \cos(\psi)\sin(\psi)\left(\sigma_{\rho\rho} - \sigma_{\psi\psi}\right) \neq 0, \tag{A.6}$$

$$\sigma_{xz} = \sigma_{yz} = \sigma_{zz} = 0.$$

In Voigt notation this gives:

$$\sigma_1 + \sigma_2 = \sigma_{\rho\rho} + \sigma_{\psi\psi} = \frac{2}{1 - (R_2/R_1)^2}\left( p_{1\rho} + \frac{R_2^2}{R_1^2}p_{2\rho}\right). \tag{A.7}$$

Let us underline, that Eq.(A.7) becomes invalid at $R_2/R_1 \to 1$ and $p_{1\rho} \neq -p_{2\rho}$, since the stress appeared in the case is too high to use the linear decoupling approximation for elastic problem. Eq.(A.7) is valid at $R_1 - R_2 >> \sqrt{\delta}$, since the surface tension coefficient itself is a collective characteristic of a surface between two phases, but not of the one or several monolayers coating the substrate or porous.[26] Moreover, at $1 - (R_2/R_1)^2 << 1$ the stress (A.7) becomes high enough for dislocations appearance. The condition $|\sigma(R_1, R_2)| \geq \sigma_{max}$ immediately leads to the drastic relaxation of stress[28, 36] and therefore avoid unphysical divergence in Eq.(A.7) at $R_1 - R_2 \to 0$. In order to take into account aforementioned effects and hereinafter we supposed that relaxation of stress is exponential, e.g. $\sigma_{1,2} = \sigma_{1,2}\exp\left(-|\sigma_{1,2}|/\sigma_{max}\right)$. Thus we modified (A.7) as following:

$$\sigma_{1,2}(R_1, R_2) = \frac{2}{1 - (R_2/R_1)^2} \left( p_{1\rho} + \frac{R_2^2}{R_1^2} p_{2\rho} \right) \exp\left( -\frac{\Delta R}{R_1 - R_2} \right). \tag{A.8}$$

Where $\Delta R$ is the critical thickness, below which the stress relaxation occurs. Rigorously speaking it should be found self-consistently from the condition $|\sigma(R_1, R_2)| \leq \sigma_{max}$ at $R_1 - R_2 \leq \Delta R$. In particular case $\mu_1 = \mu_2$, one obtains that $\Delta R \cong \mu_1/\sigma_{max}$.

The anzats of solutions (A.7-8) into the free energy (A.1) for the polarization dependent part of the free energy y gives the expression:

$$F = \left( a_1 - 2Q_{12}\sigma_{1,2}(R_1, R_2) \right) P_3^2 + a_{11} P_3^4 + a_{111} P_3^6 \tag{A.9}$$

The minimization of free energy (A.8) on the polarization components $\partial F/\partial P_3 = E_0$ gives the equation of state.

Note, that the renormalization of coefficient $a_1* = \left( a_1 - 2Q_{12} p_{ef} \right)$ for a nanotube differs from the one $a_1* = \left( a_1 - (Q_{11} + 2Q_{12}) p \right)$ obtained for a spherical nanoparticle recently.[24] Both results are clear owing to the fact that $\sigma_{1,2} \neq 0$ and $\sigma_3 = 0$ for a tube, whereas $\sigma_1 = \sigma_2 = \sigma_3 = -p$ for a sphere. Also let us underline that we do not take into account possible stress relaxation caused by dislocations and disclinations. This approach used by many authors (see e.g. Refs. 17, 24) is valid under the conditions discussed elsewhere.[37] Let us underline, that the surface tension does not affect the quartic term, in contrast to the films.

For the case when the radial pressure is caused by surface tension: $p_{1\rho} = -\mu_1/R_1$ (under the sidewall with positive curvature $R_1$ and outer normal $\mathbf{n} = (1,0,0)$) and $p_{2\rho} = -\mu_2/R_2$ (under the surface with negative curvature $-R_2$ and outer normal $\mathbf{n} = (-1,0,0)$) for the case when the radial pressure is caused by surface energy (see Fig.1).

## Appendix B

Let us consider the depolarization field distribution for the case of nanotube with arbitrary polarization distribution in the ambient conditions. In the equilibrium the perfect screening can be achieved so that there will be no electric field outside the particle.

The field distribution can be obtained on the basis of the electrostatic Poisson's equation for the electric potential $\varphi$:

$$\Delta\varphi(\rho, \psi, z) = 4\pi \, \mathrm{div}\, \mathbf{P}(\rho, \psi, z) \tag{B.1}$$

Here $\mathbf{P}(\rho, \psi, z) = (0, 0, P_Z)$ is the given z-component polarization distribution inside the nanotube, which has the cylindrical symmetry: $\Delta = \frac{\partial^2}{\partial z^2} + \frac{1}{\rho} \frac{\partial}{\partial \rho} \rho \frac{\partial}{\partial \rho} + \frac{1}{\rho^2} \frac{\partial^2}{\partial \psi^2}$. The boundary conditions on the particle surface has the view:

$$\varphi(\rho = R_{1,2}) = 0, \quad \varphi\left(z = \pm\frac{h}{2}\right) = 0. \tag{B.2}$$

Here $R_{1,2}$ and $h$ is the cylinder radii and height respectively. The boundary conditions (B.2) corresponds to the short-circuit ones proposed by Kretschmer and Binder[21] for a film.

The system (B.1), (B.2) can be solved by means of the separation of variables method. Since for the system of cylindrical symmetry eigen-functions of Laplace operator $\Delta$ are the Bessel functions one can find the potential $\varphi$ in the form of series:

$$\varphi(\rho, \psi, z) = \sum_{n,m=0}^{\infty} C_{mn}(z)\exp(im\psi)\left(A_{mn}J_m\left(\frac{k_{mn}\rho}{R_1}\right) + B_{mn}N_m\left(\frac{k_{mn}\rho}{R_1}\right)\right) \tag{B.3}$$

Here $J_m(x)$ and $N_m(x)$ are Bessel and Neiman functions of the m-th order respectively. The coefficients $A_{mn}$, $B_{mn}$ and eigen values $k_{mn}$ should be found from the lateral boundary conditions, Namely from the system

$$\begin{cases} A_{mn}J_m(k_{mn}) + B_{mn}N_m(k_{mn}) = 0 \\ A_{mn}J_m\left(k_{mn}\frac{R_2}{R_1}\right) + B_{mn}N_m\left(k_{mn}\frac{R_2}{R_1}\right) = 0 \end{cases}, \tag{B.4}$$

we obtained that

$$B_{mn} = -\frac{J_m(k_{mn})}{N_m(k_{mn})}A_{mn}, \tag{B.5a}$$

$$J_m\left(k_{mn}\frac{R_2}{R_1}\right)N_m(k_{mn}) - J_m(k_{mn})N_m\left(k_{mn}\frac{R_2}{R_1}\right) = 0 \quad . \tag{B.5b}$$

Functions $C_{mn}(z)$ should satisfy the following boundary problem:

$$\begin{cases} \dfrac{d^2C_{mn}(z)}{dz^2} - \left(\dfrac{k_{mn}}{R_1}\right)^2 C_{mn}(z) = \dfrac{4\pi}{M_{mn}}\displaystyle\int_0^{2\pi}d\psi\exp(-im\psi)\times \\ \qquad\qquad \times\displaystyle\int_{R_2}^{R_1}d\rho\,\rho\dfrac{\partial P_Z(\rho,\psi,z)}{\partial z}\left(J_m\left(\dfrac{k_{mn}\rho}{R_1}\right) - \dfrac{J_m(k_{mn})}{N_m(k_{mn})}N_m\left(\dfrac{k_{mn}\rho}{R_1}\right)\right) \\ C_{mn}\left(z = \pm\dfrac{h}{2}\right) = 0 \end{cases} \tag{B.6}$$

The functions norm $M_{mn} = 2\pi\displaystyle\int_{R_2}^{R_1}d\rho\,\rho\left(J_m\left(\dfrac{k_{mn}\rho}{R_1}\right) - \dfrac{J_m(k_{mn})}{N_m(k_{mn})}N_m\left(\dfrac{k_{mn}\rho}{R_1}\right)\right)^2$ was introduced[38] as well as eigen functions orthogonality were used. In accordance with the general theory of the linear second order differential equations one can find the solution of (B.6) in the form $C_{mn}(z) = \sum_{s=1}^{\infty} g_{mns}\sin\left(\dfrac{2\pi sz}{h}\right)$.

Finally we obtained that

$$\varphi(\rho,\psi,z) = \sum_{n,m=0,s=1}^{\infty} g_{mns}\exp(im\psi)\sin\left(\frac{2\pi sz}{h}\right)\left(J_m\left(\frac{k_{mn}\rho}{R_1}\right) - \frac{J_m(k_{mn})}{N_m(k_{mn})}N_m\left(\frac{k_{mn}\rho}{R_1}\right)\right),$$

$$g_{mns} = -\frac{4\pi}{\left(\frac{2\pi s}{h}\right)^2+\left(\frac{k_{mn}}{R_1}\right)^2}\int_{-h/2}^{h/2}\frac{2dz}{h}\sin\left(\frac{2\pi sz}{h}\right)\times \qquad\text{(B.7)}$$

$$\times\int_0^{2\pi}d\psi\frac{\exp(-im\psi)}{M_{mn}}\int_{R_2}^{R_1}\rho\,d\rho\frac{\partial P_Z(\rho,\psi,z)}{\partial z}\left(J_m\left(\frac{k_{mn}\rho}{R_1}\right) - \frac{J_m(k_{mn})}{N_m(k_{mn})}N_m\left(\frac{k_{mn}\rho}{R_1}\right)\right)$$

Keeping in mind (B.7) one obtains that depolarization field z-component $E_Z^d = -\partial\varphi/\partial z$ after integrating over parts acquires the form:

$$E_z^d(\rho,\psi,z) = \sum_{n,m=0,s=1}^{\infty} -\frac{4\pi\left(\frac{2\pi s}{h}\right)^2\exp(im\psi)}{\left(\frac{2\pi s}{h}\right)^2+\left(\frac{k_{mn}}{R_1}\right)^2}\cos\left(\frac{2\pi sz}{h}\right)\left(J_m\left(\frac{k_{mn}\rho}{R_1}\right) - \frac{J_m(k_{mn})}{N_m(k_{mn})}N_m\left(\frac{k_{mn}\rho}{R_1}\right)\right)P_{mns}\,,$$
$$\text{(B.8)}$$

$$P_{mns} = \int_{-h/2}^{h/2}\frac{2dz}{h}\cos\left(\frac{2\pi sz}{h}\right)\int_0^{2\pi}d\psi\frac{\exp(-im\psi)}{M_{mn}}\int_{R_2}^{R_1}\rho\,d\rho\,P_Z(\rho,\psi,z)\left(J_m\left(\frac{k_{mn}\rho}{R_1}\right) - \frac{J_m(k_{mn})}{N_m(k_{mn})}N_m\left(\frac{k_{mn}\rho}{R_1}\right)\right)$$

Note, that coefficients $P_{mns}$ coincide with the ones in polarization expansion:

$$P_Z(\rho,z) = \sum_{n,m=0,s=0}^{\infty}\exp(im\psi)\cos\left(\frac{2\pi sz}{h}\right)\left(J_m\left(\frac{k_{mn}\rho}{R_1}\right) - \frac{J_m(k_{mn})}{N_m(k_{mn})}N_m\left(\frac{k_{mn}\rho}{R_1}\right)\right)P_{mns}\,. \qquad\text{(B.9)}$$

It should be noticed that contrast to (B.8) the expansion (B.9) contains the terms with $P_{mn0}$ related to the average polarization. The difference $P_Z(\rho,\psi,z) - \int_{-h/2}^{h/2}\frac{2dz}{h}P_Z(\rho,\psi,z)$ acquires the form:

$$P_Z(\rho,\psi,z) - \int_{-h/2}^{h/2}dz\frac{2}{h}P_Z(\rho,\psi,z) = \sum_{n,m=0,s=1}^{\infty}\exp(im\psi)\cos\left(\frac{2\pi sz}{h}\right)\left(J_m\left(\frac{k_{mn}\rho}{R_1}\right) - \frac{J_m(k_{mn})}{N_m(k_{mn})}N_m\left(\frac{k_{mn}\rho}{R_1}\right)\right)P_{mns}$$
$$\text{(B.10)}$$

Let us assume the good convergence of the series in (B.9)-(B.10) and consider the particular cases.

1) In the particular case $h \ll \pi R_{1,2}$ one obtains that $\frac{1}{1+(k_{mn}h/2\pi sR_1)^2} \approx 1$ even for $s \sim 1$. Thus the approximate expression for depolarization field has the form (see (B.8)-(B.10)):

$$E_Z^d(\rho,\psi,z) \approx -4\pi\left(P_Z(\rho,\psi,z) - \int_{-h/2}^{h/2}dz\frac{2}{h}P_Z(\rho,\psi,z)\right) \qquad\text{(B.11)}$$

Note, that (B.11) is exact at $R_{1,2}\to\infty$ and $R_1 - R_2 = const$, and coincides with the one obtained for ferroelectric films[21] at $P_Z(\rho,\psi,z)\equiv P_Z(z)$.

2) In the particular case $h \gg \pi R_1$ one obtains the estimation:

$$\left|E_Z^d(\rho,\psi,z)\right| \sim 4\pi \left(\frac{2\pi}{k_{mn}}\frac{R_1}{h}\right)^2 \left(P_Z(\rho,\psi,z) - \int\limits_{-h/2}^{h/2} dz \frac{2}{h} P_Z(\rho,\psi,z)\right). \tag{B.12}$$

The interpolation for the depolarization field that contains the aforementioned particular cases (B.11)-(B.12) acquires the form:

$$E_Z^d(\rho,\psi,z) = -\eta \left(P_Z(\rho,\psi,z) - \int\limits_{-h/2}^{h/2} dz \frac{2}{h} P_Z(\rho,\psi,z)\right),$$
$$\eta = \frac{4\pi}{1+(k_{01}h/2\pi R_1)^2}. \tag{B.13}$$

where $k_{01}$ is the first root of the equation $J_0\left(k_{01}\frac{R_2}{R_1}\right)N_0(k_{01}) - J_0(k_{01})N_0\left(k_{01}\frac{R_2}{R_1}\right) = 0$ .

## Appendix C

Variation of the free energy expressions (1) - (4) yields the following Euler-Lagrange equations with the boundary conditions on the cylinder butts $z = \pm h/2$, and the sidewall surfaces $\rho = R_{1,2}$ (see e.g. Ref. 6, 22):

$$\begin{cases} \alpha_R P_Z + \beta P_Z^3 + \gamma P_Z^5 - \delta\left(\frac{\partial^2}{\partial z^2} + \frac{1}{\rho}\frac{\partial}{\partial\rho}\rho\frac{\partial}{\partial\rho} + \frac{1}{\rho^2}\frac{\partial^2}{\partial\psi^2}\right)P_Z = E_0 + E_Z^d(\rho,\psi,z), \\ \left(P_Z + \lambda_b \frac{dP_Z}{dz}\right)\bigg|_{z=h/2} = 0, \qquad \left(P_Z - \lambda_b \frac{dP_Z}{dz}\right)\bigg|_{z=-h/2} = 0, \\ \left(P_Z + \lambda_S \frac{dP_Z}{d\rho}\right)\bigg|_{\rho=R_1} = 0, \qquad \left(P_Z - \lambda_S \frac{dP_Z}{d\rho}\right)\bigg|_{\rho=R_2} = 0, \end{cases} \tag{C.1}$$

Let us find the approximate solution of the nonlinear Eq.(C.1) by using the direct variational method as proposed earlier.[22] Firstly we solved the linearized Eq.(C.1) allowing for (B.13). Under the assumption $(\lambda_S/R_1) << 1$ we obtained that

$$P_Z(\rho,\psi,z) = \sum_{n,m=0}^{\infty} \exp(im\psi)\left(J_m\left(\frac{k_{mn}\rho}{R_1}\right) - \frac{J_m(k_{mn})}{N_m(k_{mn})}N_m\left(\frac{k_{mn}\rho}{R_1}\right)\right)\frac{(E_{mn}^0 + \eta P_{mn}^0)(1-\Phi_{mn}(z))}{\alpha_R + \eta + \delta(k_{mn}/R_1)^2} \tag{C.2}$$

Here:

$$E_{mn}^0 = \int\limits_0^{2\pi} d\psi \frac{\exp(-im\psi)}{M_{mn}} \int\limits_{R_2}^{R_1} \rho d\rho \left(J_m\left(\frac{k_{mn}\rho}{R_1}\right) - \frac{J_m(k_{mn})}{N_m(k_{mn})}N_m\left(\frac{k_{mn}\rho}{R_1}\right)\right)E_0 \tag{C.3a}$$

$$P_{mn}^0 = \int\limits_0^{2\pi} d\psi \frac{\exp(-im\psi)}{M_{mn}} \int\limits_{R_2}^{R_1} \rho d\rho \left(J_m\left(\frac{k_{mn}\rho}{R_1}\right) - \frac{J_m(k_{mn})}{N_m(k_{mn})}N_m\left(\frac{k_{mn}\rho}{R_1}\right)\right)\int\limits_{-h/2}^{h/2} dz \frac{2}{h} P_Z(\rho,\psi,z) \tag{C.3b}$$

$$\Phi_{mn}(z) = \frac{ch(\xi_{mn}z)}{ch\left(\xi_{nm}\frac{h}{2}\right) + \lambda_b\xi_{nm}sh\left(\xi_{nm}\frac{h}{2}\right)}, \tag{C.3c}$$

$$\begin{cases} \xi_{mn} = \sqrt{(k_{mn}/R_1)^2 + (\alpha_R + \eta)/\delta} \\ J_m\left(k_{mn}\dfrac{R_2}{R_1}\right)N_m(k_{mn}) - J_m(k_{mn})N_m\left(k_{mn}\dfrac{R_2}{R_1}\right) = 0 \end{cases} \qquad (C.3d)$$

Here $J_m(x)$ and $N_m(x)$ are Bessel and Neiman functions of the m-th order respectively. The functions norm $M_{mn} = 2\pi \int\limits_{R_2}^{R_1} d\rho\,\rho\left(J_m\left(\dfrac{k_{mn}\rho}{R_1}\right) - \dfrac{J_m(k_{mn})}{N_m(k_{mn})}N_m\left(\dfrac{k_{mn}\rho}{R_1}\right)\right)^2$ was introduced.[38]

For the infinite nanotube ($h \to \infty$ values $\eta \to 0$), one obtains that paraelectric dielectric susceptibility

$\chi_{33} = \dfrac{d\,P_Z}{d\,E_0}$ diverges under the conditions $\alpha_R + \delta(k_{mn}/R_1)^2 = 0$. The lowest root $k_{mn}(R_1, R_2)$

corresponds to the case $m = 0$, $n = 1$. In fact the roots are tabulated and depend on the ratio $R_2/R_1$

only (see Fig.1C).

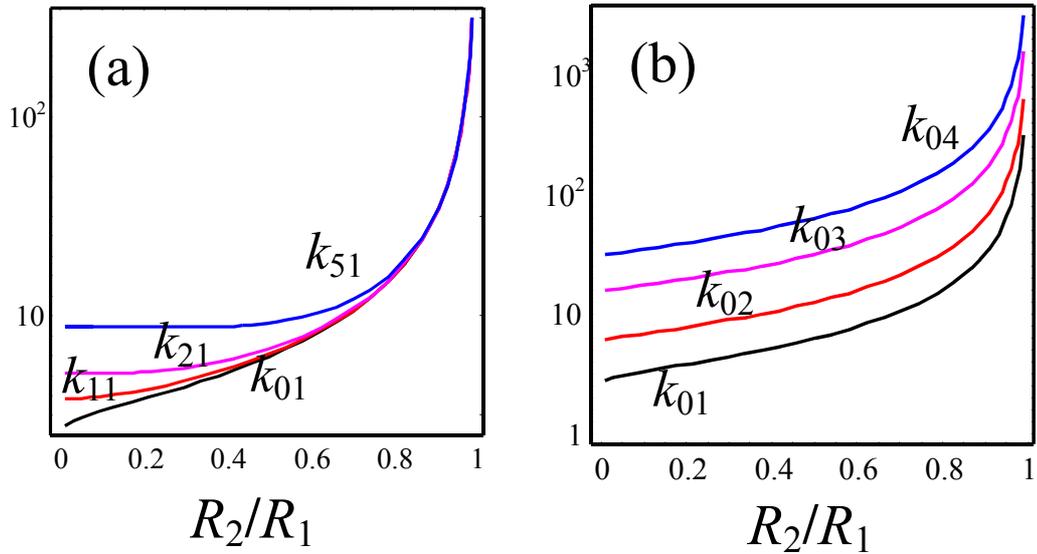

Fig. 1C. (Color online) The roots $k_{mn}(R_1, R_2)$ vs. the ratio $R_1/R_2$.

Thus paraelectric phase loses its stability under the condition $\alpha_R + \delta(k_{01}/R_1)^2 = 0$, which

immediately gives the parametric expression for the transition temperature $T_{CR}(R_1, R_2)$:

$$\begin{cases} T_{CR}(R_1, R_2) = T_C + \dfrac{2Q_{12}\exp\left(-\dfrac{\Delta R}{R_1 - R_2}\right)}{\alpha_T\left(1 - (R_2/R_1)^2\right)}\left(p_1 + \left(\dfrac{R_2^2}{R_1^2}\right)p_2\right) - \delta\dfrac{k_{01}^2(R_1, R_2)}{\alpha_T R_1^2}, \\ J_0\left(k_{01}\dfrac{R_2}{R_1}\right)N_0(k_{01}) - J_0(k_{01})N_0\left(k_{01}\dfrac{R_2}{R_1}\right) = 0 \end{cases} \qquad (C.4)$$

The paraelectric polarization distribution inside the infinite nanotube acquires the form (see (C.2)):

$$P_Z(\rho,\psi) = \sum_{n,m=0}^{\infty} \exp(im\psi)\left(J_m\left(\frac{k_{mn}\rho}{R_1}\right) - \frac{J_m(k_{mn})}{N_m(k_{mn})}N_m\left(\frac{k_{mn}\rho}{R_1}\right)\right)\frac{E_{mn}^0}{\alpha_R + \delta(k_{mn}/R_1)^2} \quad \text{(C.5)}$$

The polarization distribution in the ferroelectric phase should be found by direct variational method, namely the anzats $\dfrac{E_{mn}^0}{\alpha_R + \delta(k_{mn}/R_1)^2} \to \dfrac{P_{mn}^V}{M_{mn}}$ should be used. When substituting (C.5) into the free energy $\Delta G = \Delta G_V + \Delta G_S$ (see Eqs.(1) and (4)) and integrating over nanoparticle volume, we obtained the free energy with renormalized coefficients. For a single-domain particle it has the form:

$$\Delta G = \frac{\alpha_R(T) + \delta(k_{01}/R_1)^2}{2}P_Z^2 + \frac{\beta}{4}P_Z^4 + \frac{\gamma}{6}P_Z^6 - P_Z E_0 \quad \text{(C.6)}$$

From (C.6) one immediately obtains the average polarization, coercive field etc.

## Appendix D

In particular case of effective point charge representing the probe:

$$t_{13}(R_1,R_2,\gamma,\nu) \approx -\left(\begin{array}{l}\dfrac{1+2\gamma}{(1+\gamma)^2}\cdot\left(\dfrac{R_1}{R_1 + D_{33}(\gamma)\cdot d_Q} - \dfrac{R_2}{R_2 + D_{33}(\gamma)\cdot d_Q}\right)- \\[4mm] -\dfrac{2(1+\nu)}{1+\gamma}\cdot\left(\dfrac{R_1}{R_1 + D_{13}(\gamma)\cdot d_Q} - \dfrac{R_2}{R_2 + D_{13}(\gamma)\cdot d_Q}\right)\end{array}\right) \quad \text{(D.1)}$$

$$t_{51}(R_1,R_2,\gamma) \approx \frac{\gamma^2}{(1+\gamma)^2}\cdot\left(\frac{R_1^2}{2\dfrac{\gamma^2 d_Q^2}{(1+\gamma)^2} + D_{51}(\gamma)R_1 d_Q + R_1^2} - \frac{R_2^2}{2\dfrac{\gamma^2 d_Q^2}{(1+\gamma)^2} + D_{51}(\gamma)R_2 d_Q + R_2^2}\right) \quad \text{(D.2)}$$

$$t_{33}(R_1,R_2,\gamma) \approx \frac{1+2\gamma}{(1+\gamma)^2}\cdot\left(\frac{R_1}{R_1 + D_{33}(\gamma)\cdot d_Q} - \frac{R_2}{R_2 + D_{33}(\gamma)\cdot d_Q}\right) \quad \text{(D.3)}$$

Here $d_Q = \varepsilon_e R_0/\sqrt{\varepsilon_{11}\varepsilon_{33}}$ is the distance between the sample surface and the effective point charge representing the probe ($R_0$ is the probe apex curvature, $\varepsilon_e$ is ambient permittivity, $d_Q \sim 10\,nm$). Usually $d_Q$ could be regarded fitting parameter, functions $D_{ij}(\gamma)$ are listed in Ref. 32.